\documentclass[10pt]{revtex4}
\usepackage{amssymb}
\usepackage{latexsym}
\usepackage{epsfig}
\usepackage{float}
\usepackage{graphicx,epsfig,color}
\usepackage{graphicx}
\usepackage{subfigure}
\usepackage{amsmath}
\usepackage[T1]{fontenc}
\begin{document}
\title{Lifshitz and hyperscaling violated Yang-Mills-dilaton black holes}
\author{Fatemeh Naeimipour$^{1}$\footnote{sara.naeimipour1367@gmail.com}, Behrouz Mirza$^{1}$\footnote{b.mirza@iut.ac.ir}, and Soudabe Nasirimoghadam $^{2}$\footnote{snasirimoghadam@sirjantech.ac.ir}}
\affiliation{$^1$ Department of Physics, Isfahan University of Technology, Isfahan, 84156-83111, Iran\\
$^2$ Department of Physics, Sirjan University of Technology, Sirjan, 78137, Iran}
\begin{abstract}
We obtain two types of Yang-Mills-dilaton black hole solutions in both Lifshitz and hyperscaling violation spacetimes. We must consider at least three Yang-Mills gauge fields that interact with a scalar field and either $SO(n)$ or $SO(n-1,1)$ gauge symmetry groups, where $n+1$ denotes the dimension of the spacetime. They lead to the spherical and hyperbolic solutions. The obtained solutions in the hyperscaling violation spacetime fall into two categories for $z\neq n-3-\frac{n-5}{n-1}\theta$ and $z= n-3-\frac{n-5}{n-1}\theta$, where $\theta=0$ represents the Lifshitz spacetime. In order to have a real asymptotic behavior for the hyperscaling violated black hole solutions, we should consider a negative value for the hyperscaling violation parameter as $\theta<0$. We also evaluate the thermodynamic quantities of the mentioned black holes and probe their thermal stability in the grand canonical ensemble. For $z\geq2$, the hyperscaling violated solutions are not thermally stable for $z\leq n-3-\frac{n-5}{n-1}\theta$, while they are stable for large $r_{+}$ with $z>n-3-\frac{n-5}{n-1}\theta$. We also check out the critical behavior of the obtained black holes and obtain a Smarr relation for the solutions. The results also announce of a first-order small-large phase transition for both black holes in the case $T>T_{C}$. 
\end{abstract}


\maketitle

\section{Introduction}\label{intro}
AdS/CFT correspondence plays an important role in the study of the strongly coupled systems. It relates these systems in quantum field theory to the weakly coupled classical relativistic gravitational ones in anti-de Sitter (AdS) spacetime \cite{Malda1,Witt,Gubser1}. This correspondence is a useful tool to analyze the condensed matter
systems. It was shown that some properties of the strongly coupled superconductors are descriptive by their dual gravitational spacetimes \cite{Hart0,Hart1,Hart11,Hart33}. 
There are some motivations to study the nonrelativistic version of AdS/CFT in order to gain information about the puzzles in unconventional
condensed matter physics \cite{Nick,Hoyo,Nish}. Some 
real condensed matter systems do not have relativistic symmetry, and they are described at their critical temperatures by the nonrelativistic conformal field theories \cite{Hartno,Herzog}. The other is related to the fermions at unitarity which can be discovered experimentally by cold atoms. The interactions of these fermions are adapted to provide a scale-invariant but nonrelativistic system. Another reason originates from the high-temperature superconductors problem in modern condensed matter physics. AdS/CFT has been successful to investigate all of them by the nonrelativistic general relativity. Spacetimes with the 
so-called Schr$\rm\ddot{o}$dinger group are the gravity
duals for the nonrelativistic scale-invariant systems with Galilean symmetry \cite{Son,Bala}. Lifshitz spacetimes are the other gravity dual candidates which have no Galilean symmetry \cite{kach}. This spacetime is defined by the metric 
\begin{eqnarray}\label{met1}
ds^2=-\frac{r^{2z}}{L^{2z}}dt^2+\frac{L^2}{r^2}dr^2+r^2 d \vec{x}_{n-1}^{2},
\end{eqnarray}
where $z>0$ and $L$ are, respectively the dynamical critical exponent and the AdS radius and $n+1$ counts the spacetime dimensions. This metric is not conformally invariant, while it shows an anisotropic scaling symmetry 
\begin{eqnarray}\label{tran1}
t\rightarrow\lambda^z t\,\,\,\,,\,\,\, \vec{x}\rightarrow\lambda  \vec{x}\,\,\,\,\,,\,\,\,r\rightarrow\lambda^{-1} r.
\end{eqnarray}
This kind of spacetime can be a dual with nonrelativistic conformal field theory describing multicritical points in certain magnetic materials and liquid crystals \cite{kach}. For $z=1$, the metric reduces to the usual one with the relativistic isotropic scale invariance. The phase transitions of many condensed matter systems are governed by fixed points with the above scaling in Eqs. \eqref{tran1}. Also, with the above anisotropy, the specific heat scales at low temperature as $C_{V}\sim T^{(n-1)/z}$. As the specific heat of the Fermi liquids obeys from the linear relation $C_{V}\sim T$, so Lifshitz scaling theories are good candidates in order to analyze Fermi liquids for $z=n-1$ \cite{Kok}. Using AdS/CFT correspondence, some properties such as viscosity \cite{Pang} and conductivity \cite{Sin,Lemos1,Qian,Roych} have been studied for systems with Lifshitz symmetry. In Ref.\cite{B3}, the effects of the Lifshitz dynamical exponent and the Weyl coupling on the holographic superconductors is investigated. Entanglement entropy was also used to investigate holographic superconductor phase transitions related to Lifshitz black hole background \cite{B1}.\\
Einstein garvity with a cosmological constant cannot have Lifshitz solutions except for $z=1$. For $z\neq 1$, Einstein's equations may allow Lifshitz solutions only by adding some higher-curvature tensors or matter sources to this gravity. Studies of some modified gravities such as Lovelock and quasitopological ones in the Lifshitz spacetime are in Refs. \cite{Deh0,Deh2,Deh3,Bazr1}. In Refs.  \cite{Maeda,Lee,Alva,kach,Taylor,Pang}, the Lifshitz black hole solutions with matter sources such as Brans-Dicke scalars, nonlinear electrodynamic
theories and Proca fields have been investigated. Some investigations of the matter sources with massive gauge fields are in Refs. \cite{Pang,Mann1,Ber,Bal}. They illustrate that it is not possible to obtain an exact Lifshitz solution for their theories. To solve this problem, Tarrio considered a dilaton field instead of the massive gauge fields \cite{Tarrio}, which led to an analytic Lifshitz black hole solution. Since the dilaton field plays as a requirement of the
string theory, so it appears at the low-energy limit of this theory and we should include in our study. Lifshitz black holes with a dilaton field have been probed in Refs. \cite{Bert,Zan1}. Now, in continuation of Ref. \cite{Tarrio}, we would like to obtain the exact Lifshitz-dilaton black hole solutions in the presence of an interesting matter source, i.e., non-Abelian Yang-Mills gauge theory.\\
Recently, some numeric solutions of the Lifshitz black holes in Einstein-Yang-Mills theory
have been obtained \cite{Dev}. In our paper, we aim to use the Wu-Yang ansatz \cite{Wu} 
and obtain the exact solutions. Some black hole solutions in the presence of the Yang-Mills theory are in Refs. \cite{Yang1,Yang2,Yang3}. A study of the Einstein-Yang-Mills-dilaton black hole in higher dimensions is in Ref. \cite{Yang4}. Topological black holes in $(n+1)$-dimensional Einstein and Gauss-Bonnet gravities have been probed, respectively, in Refs. \cite{Yang5,Yang6}.\\
Magnetic monopoles and their condensation are necessary in order to describe quark confinement with a dual superconductor picture. For this purpose, the Abelian projection method \cite{Hoo} explicitly
breaks both the local gauge symmetry and the global color
symmetry by partial gauge fixing. A new gauge-invariant way in SU(N) Yang-Mills theory has been successful
in order to introduce gauge-invariant non-Abelian magnetic monopoles \cite{Kondo}. These non-Abelian magnetic monopoles have a dominant
contribution for confinement of fundamental quarks in SU(3) Yang-Mills theory. There are also some topological protected
quantum computations in which non-Abelian
excitations such as Majorana fermions are used \cite{Ivanov,Kitaev,Tewari,Nayak}. Non-Abelian toplogical superconductors can be determined by these fermions bound in the quantized vortices \cite{Jia}. On the other hand, as the spin currents of the ferromagnets correspond to the $SU(2)$ gauge fields in the dual gravitational
theory \cite{Yoko}, one can obtain novel perspectives for the condensed matter systems dual to the non-Abelian Yang-Mills gauge theories. Our main motivation to study Lifshitz-Yang-Mills black holes was some of the above-mentioned non-Abelian nonrelativistic physical systems with strongly coupled interactions. \\
Recently, a new metric has been introduced that not only is not scale invariant, but also has a hyperscaling violation factor  \cite{hyper1,hyper2,hyper3}. In the AdS/CFT correspondence, the hyperscaling violation in the field theory originates from the distance noninvariancy of the dual metric under the scaling. This metric is defined as 
\begin{eqnarray}\label{met2} 
ds^2=r^{-\frac{2\theta}{n-1}}\bigg(-\frac{r^{2z}}{L^{2z}}dt^2+\frac{L^2 }{r^2}dr^2+r^2 d\vec{x}_{n-1}^2\bigg),
\end{eqnarray}
where $\theta$ is referred to the hyperscaling violating exponent and can decrease the dimension of the theory. It is spatially homogeneous and covariant under the scale transformations \eqref{tran1}. For $\theta=0$, this metric reestablishes the Lifshitz spacetime. For the special case $\theta=n-2$, the related spacetime can be a dual theory for the Fermi liquid \cite{hyper3}. Some solutions of the hyperscaling violation spacetimes have been found in Refs. \cite{Charm,Gath,Ghod,Ganj}. In the second part of this paper, we aim to extend our study and obtain the dilaton Yang-Mills black hole solutions with a hyperscaling violation.\\
We divide the organization of this paper into two parts: Lifshitz and hyperscaling violated Yang-Mills-dilaton black holes. For the Lifshitz black holes, we first consider the Yang-Mills theory with a dilaton field, then obtain the Lifshitz solutions, and study the related physical properties in Sec. \ref{Field}. In Sec. \ref{thermo1}, we study the thermodynamic behaviors of the obtained Lifshitz solutions such as thermal stability. We also study the critical behavior of the Lifshitz Yang-Mills-dilaton solutions in Sec. \ref{PV1}. In the second part of this paper, we go to the Yang-Mills-dilaton black holes with a hyperscaling violation parameter. In Sec.\ref{field2}, we obtain the related solutions and then investigate the thermodynamic and critical behavior of the obtained solutions in respectively, Secs.\ref{thermo2} and \ref{PV2}. Last, a conclusion of the whole paper is in Sec.\ref{result}.     
\section{The main structure of the Lifshitz Yang-Mills-dilaton black hole}\label{Field}
\begin{figure}
\centering
\subfigure[$e_{3}=2$, $n=4$]{\includegraphics[scale=0.27]{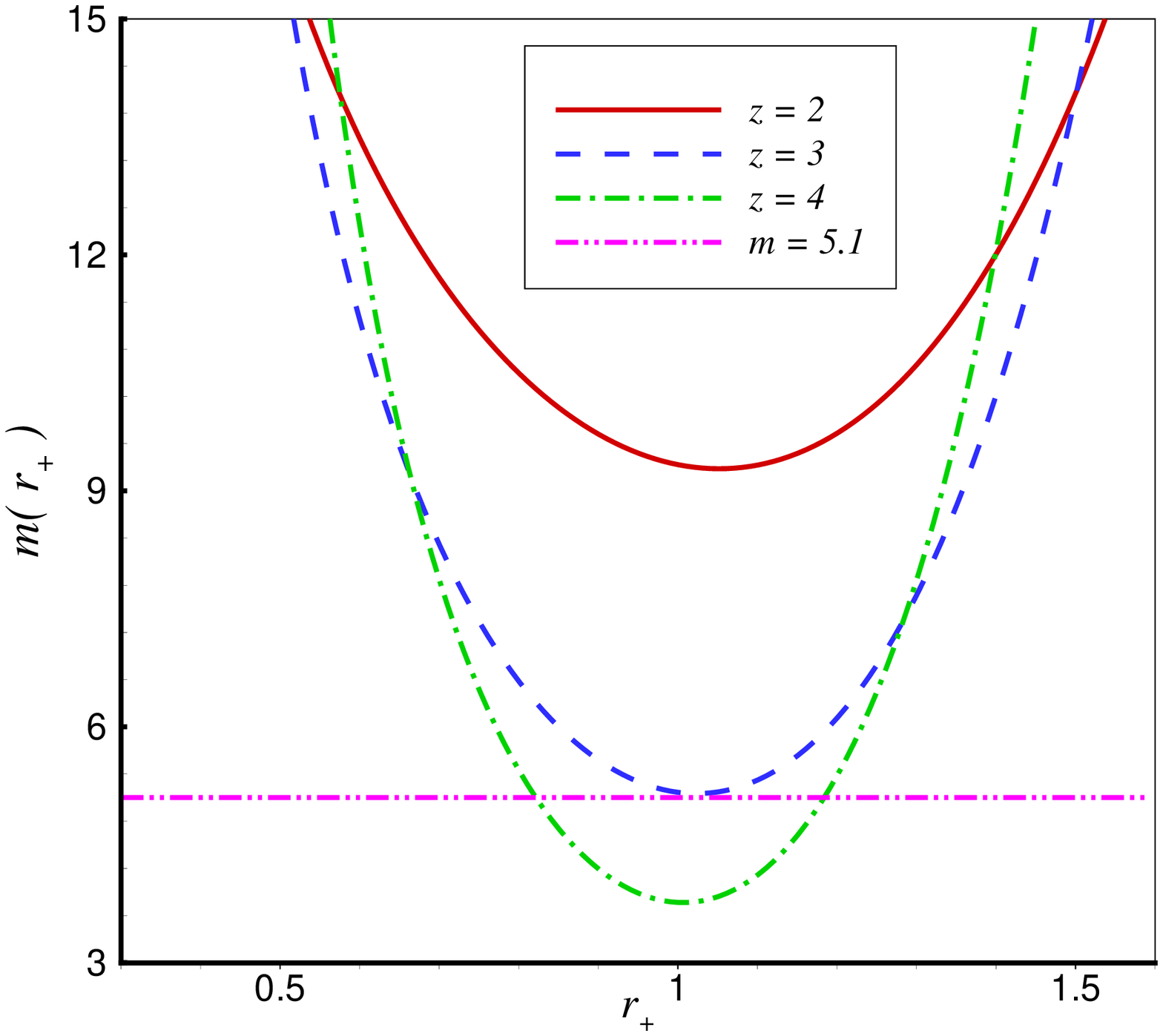}\label{Fig1a}}\hspace*{.2cm}
\subfigure[$z=2$, $n=4$]{\includegraphics[scale=0.27]{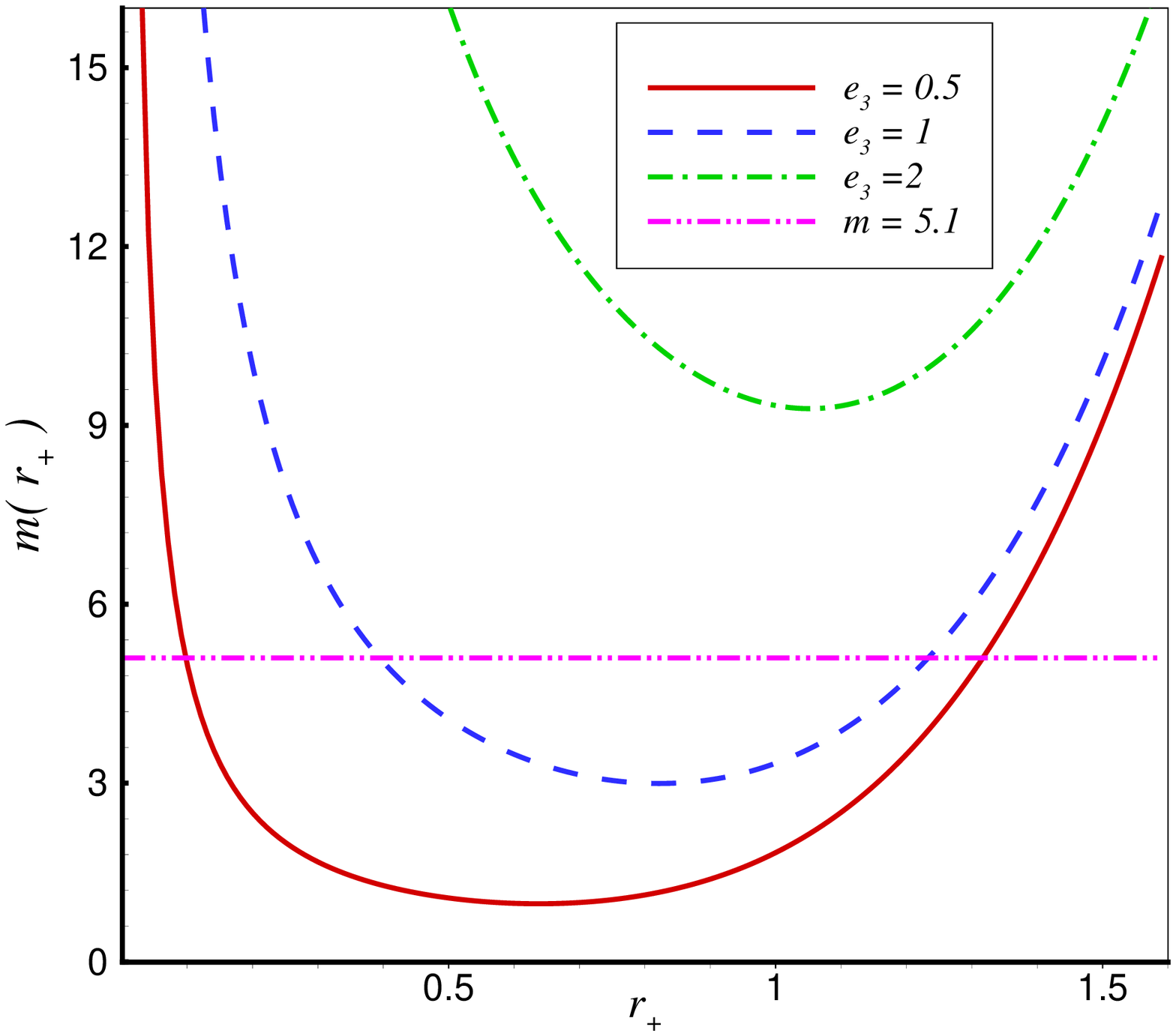}\label{Fig1b}}\hspace*{.2cm}
\subfigure[$e_{3}=1$, $z=3$]{\includegraphics[scale=0.27]{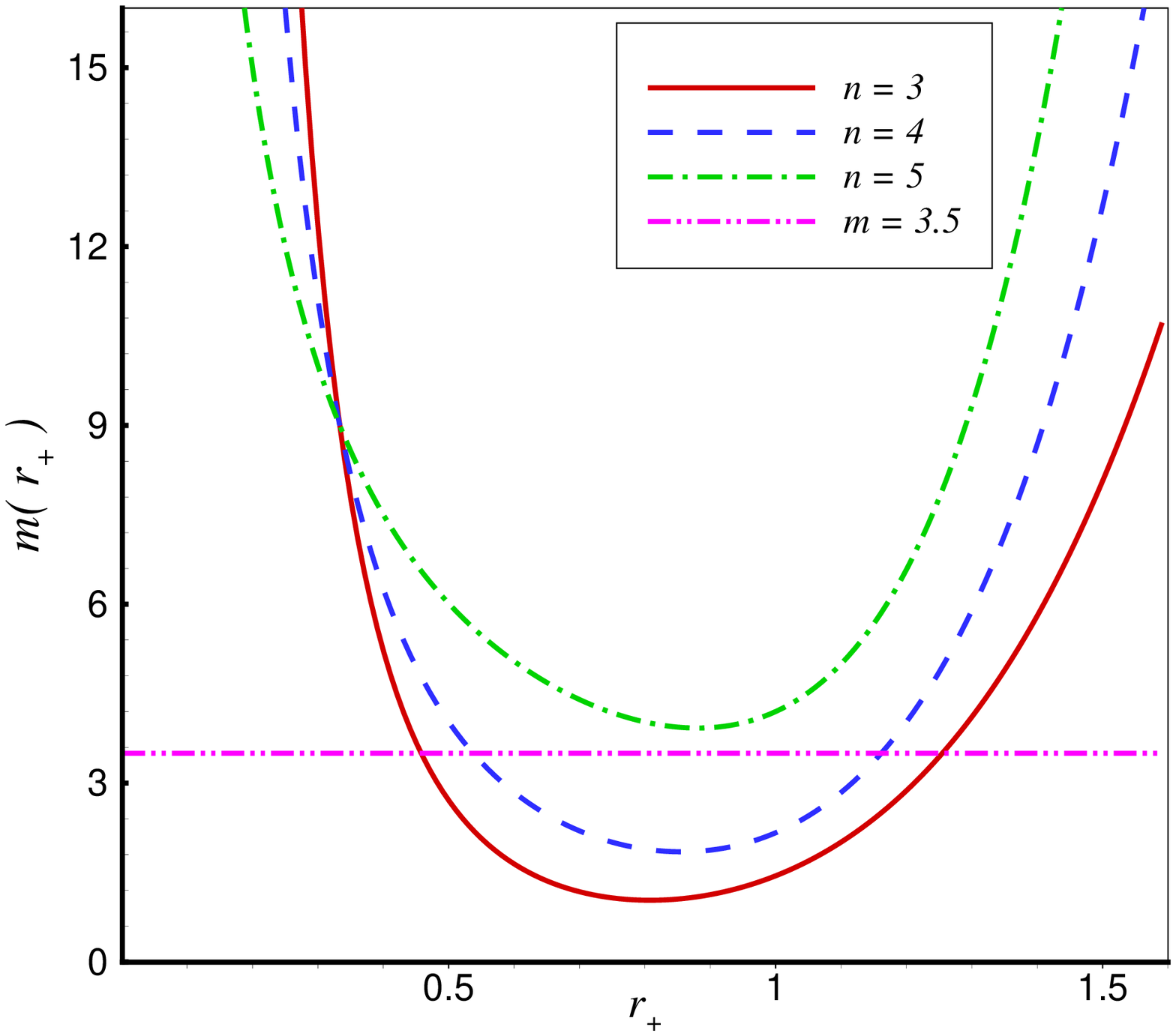}\label{Fig1c}}\caption{The function $m(r_{+})$ with respect to $r_{+}$ for $s=1$ and $L=1$.}\label{Fig1}
\end{figure}

In this section, we aim to obtain the Lifshitz-dilaton black hole solutions in the presence of the Yang-Mills theory. For this purpose, we consider three non-Abelian gauge fields $F_{i}$ $(i=1,2,3)$ with gauge groups $\mathcal{G}$:
\begin{eqnarray}
F_{i\mu\nu}^{(a)}=\partial_{\mu}A_{i\nu}^{(a)}-\partial_{\nu}A_{i\mu}^{(a)}+\frac{1}{e_{i}}C^{a}_{bc}A_{i\mu}^{(b)}A_{i\nu}^{(c)},
\end{eqnarray}
where $e_{i}$'s $(e_{1},e_{2},e_{3})$ are the coupling constants for each $F_{i}$'s $(F_{1},F_{2},F_{3})$ and $A_{i\mu}$'s refer to the gauge potentials. 
We start our theory with the $(n+1)$-dimensional action 
\begin{eqnarray}\label{action}
S=\frac{1}{16\pi}\int{d^{n+1}x \sqrt{-g}\bigg(R-V(\Phi)-\frac{4}{n-1}\partial_{\mu}\Phi\partial^{\mu}\Phi-\sum_{i=1}^{3} e^{-4\xi_{i}\Phi/(n-1)}F_{i}^2\bigg)},
\end{eqnarray}
where $\Phi$ displays the dilaton field and $V(\Phi)$ is a potential for this field. $\xi_{i}$ is a constant which measures the coupling strength of $\Phi$ and $F_{i}^2$,  
\begin{eqnarray}
F_{i}^2=\gamma_{ab}F_{i\mu\nu}^{(a)}F_{i}^{(b)\mu\nu}, 
\end{eqnarray}
where
\begin{eqnarray}
\gamma_{ab}\equiv-\frac{\Gamma_{ab}}{|\rm det \Gamma_{ab}|^{1/N}} \,\,\,\, ,\,\,\, \Gamma_{ab}=C_{ad}^{c}C_{bc}^{d},
\end{eqnarray}
that $N$ represents the parameters of each groups $\mathcal{G}_{i}$ and $C_{bc}^{a}$'s are the structure constants of the groups.
By variation of the action \eqref{action} with respect to $g_{\mu\nu}$, $A_{i\nu}^{(a)}$ and $\Phi$, the field equations of motion are obtained as follows
\begin{eqnarray}\label{eq1}
R_{\mu\nu}=\frac{2}{n-1}\bigg[2\partial_{\mu}\Phi \partial_{\nu}\Phi+\frac{V(\Phi)}{2} g_{\mu\nu}\bigg]+2\sum_{i=1}^{3}e^{-4\xi_{i}\Phi/(n-1)}\bigg[\gamma_{ab}F_{i\mu}^{(a)\lambda}F_{i\nu\lambda}^{(b)}-\frac{1}{2(n-1)}F_{i}^2g_{\mu\nu}\bigg],
\end{eqnarray}
\begin{eqnarray}\label{eq2}
D_{\nu}(e^{-4\xi_{i}\Phi/(n-1)}F_{i}^{(a)\mu\nu})=\frac{1}{e_{i}}e^{-4\xi_{i}\Phi/(n-1)}C_{bc}^{a}A_{i\nu}^{(b)}F_{i}^{(c)\nu\mu},
\end{eqnarray}
\begin{eqnarray}\label{eq3}
\nabla^{2}\Phi-\frac{n-1}{8}\frac{dV(\Phi)}{d\Phi}+\sum_{i=1}^{3}\frac{\xi_{i}}{2} e^{-4\xi_{i}\Phi/(n-1)}F_{i}^2=0.
\end{eqnarray}
If we denote the coordinates of the groups $\mathcal{G}_{i}$ as below
\begin{eqnarray}\label{coor}
x_{1}&=&\frac{r}{\sqrt{k}}\, \mathrm{sin}(\sqrt{k}\,\theta)\,\Pi_{j=1}^{n-2}\,\mathrm{sin}\,\phi_{j},\nonumber\\
x_{l}&=&\frac{r}{\sqrt{k}}\, \mathrm{sin}(\sqrt{k}\,\theta)\,\mathrm{cos}\,\phi_{n-l}\,\Pi_{j=1}^{n-l-1}\,\mathrm{sin}(\phi_{j})\,\,\,\,,\,\,\,l=2, .\,.\,., n-1\nonumber\\
x_{n}&=& r\, \mathrm{cos}\,(\sqrt{k}\,\theta),
\end{eqnarray}
then we can attain the gauge potentials by the Wu-Yang ansatz \cite{Wu}:
\begin{eqnarray}\label{poten}
A_{i}^{(a)}&=&\frac{e_{i}}{r^2}(x_{l}dx_{n}-x_{n}dx_{l})\,\,\,\mathrm{for}\,\,\, a=l=1,.\,.\,.,n-1,\nonumber\\
A_{i}^{(b)}&=&\frac{e_{i}}{r^2}(x_{l}dx_{j}-x_{j}dx_{l})\,\,\, \mathrm{for}\,\,\,\,b=n,.\,.\,.,n(n-1)/2,\,\,\,l=1,.\,.\,.,n-2,\,\,\,j=2,.\,.\,.,n-1,\mathrm{and}\,l<j,
\end{eqnarray}
which have the Lie algebra of $SO(n)$ and $SO(n-1,1)$ gauge groups.
To obtain the Lifshitz solutions, we consider the following metric: 
\begin{eqnarray}\label{metric}
ds^2=-\frac{r^{2z}}{L^{2z}}f(r)dt^2+\frac{L^2}{r^2f(r)}dr^2+r^2[d\theta^2+k^{-1}\mathrm{sin}^{2} (\sqrt{k}\theta)d\Omega_{k,n-2}^2],
\end{eqnarray}
where $d\Omega_{k,n-2}^2$ denotes the metric of a unit $(n-2)$-sphere with constant curvatures $k=-1,1$. To be clear, we have mentioned some examples of the gauge groups in the appendix \eqref{app}. We consider $V(\Phi)=2\Lambda$, where $\Lambda$ is the cosmological constant. Now, if we substitute the gauge potentials \eqref{poten} and the metric \eqref{metric} in Eqs. \eqref{eq1}\,-\eqref{eq3}, then the Lifshitz-dilaton solutions of the Yang-Mills theory read as
\begin{eqnarray}\label{phi1}
\Phi(r)=\frac{n-1}{2}\sqrt{z-1}\mathrm{ln}\bigg(\frac{r}{s}\bigg)
\end{eqnarray}
and
\begin{eqnarray}\label{ff}
f(r)&=&1+\frac{kL^2(n-2)}{z (z+n-3)r^2}-\frac{m}{r^{z+n-1}}+\left\{
\begin{array}{ll}
$$\frac{(n-2)L^2 e_{3}^2}{(z-n+3)r^{2z+2}}s^{2(z-1)}$$,\quad\quad\quad \quad  \ {\mathrm{for}\,\, z\neq n-3, }\quad &  \\ \\
$$-\frac{(n-2)L^2 e_{3}^2}{r^{2z+2}}\mathrm{ln}(\frac{r}{r_{0}})s^{2(z-1)}$$,\quad\quad\quad  \ {\mathrm{for}\,\, z=n-3,}\quad &
\end{array}
\right.
\end{eqnarray}
where $s$ and $m$ are the constants of integration and $m$ is related to the mass of the black hole. We use $r_{0}$ as a necessary parameter to provide a dimensionless argument for the logarithmic
term. We set $r_{0}=1$ for simplicity. In order to obtain the above asymptotic Lifshitz solutions, we have fixed the parameters $\xi_{i}$($i=1,2,3$), $e_{1}^2$, $e_{2}^2$, and $\Lambda$ as below:
\begin{eqnarray}\label{con1}
\xi_{1}&=&-\frac{2}{\sqrt{z-1}},\,\,\xi_{2}=-\frac{1}{\sqrt{z-1}}\,\,,\,\,\xi_{3}=\sqrt{z-1}\,\,,\,\,\nonumber\\
e_{1}^2&=-&\frac{2(z-1)\Lambda}{(n-1)(n-2)(z+1)}s^{4}\,\,\,\,,\,\,\,
e_{2}^2=k\frac{z-1}{z}s^{2},\nonumber\\
\Lambda&=&-\frac{(n-1)(z+1)(z+n-1)}{4L^2}.
\end{eqnarray}
The results represent that, for $n=3$, the Lifshitz-Yang-Mills dilaton solutions are the same as the ones in Maxwell theory. So, we can deduce that there is an equivalence between the Lifshitz-Yang-Mills-dilaton solutions of $SO(3)$
and $SO(2,1)$ gauge groups and a set of topological solutions with $k=+1,-1$ in the linear Maxwell theory. For $n\neq 3$, we could have been able to reach to a new class of Lifshitz-Yang-Mills solutions which are different from the ones in the Maxwell theory. The obtained solutions in Eq. \eqref{ff} are divided into two parts: for $z\neq n-3$ and $z=n-3$.\\
The parameters $\xi_{1}$ and $\xi_{2}$ diverge at $z=1$. However, it should be noted that the parameters $\xi_{i}$'s are multiplied by the dilaton field $\Phi(r)=(n-1)\sqrt{z-1}\,\mathrm{ln} (r/s)/2$ in the field equations \eqref{eq1} -\eqref{eq3}, so the coefficient $\sqrt{z-1}$ in the denominator cancels out and there is no ambiguity. At $z=1$, both $e_{1}$ and $e_{2}$ in Eq. \eqref{con1} are equal to zero, and, therefore, the tensor fields $F_{1}$ and $F_{2}$ have zero values. The factor $e^{-4\xi_{i}\Phi/(n-1)}$ is multiplied to the zero tensor fields $F_{1}$ and $F_{2}$, and the result is zero. So, at $z=1$, we recover the Einstein-Yang-Mills theory with no ambiguity: 
\begin{eqnarray}\label{fz=1}
f(r)&=&1+\frac{kL^2}{r^2}-\frac{m}{r^{n}}+\left\{
\begin{array}{ll}
$$-\frac{(n-2)L^2 e_{3}^2}{(n-4)r^{4}}
$$,\quad\quad\quad \quad  \ {\mathrm{for}\,\, n\neq 4, }\quad &  \\ \\
$$-\frac{(n-2)L^2 e_{3}^2}{r^{4}}\mathrm{ln}\bigg(\frac{r}{r_{0}}\bigg)$$,\quad\quad\quad  \ {\mathrm{for}\,\, n=4.}\quad &
\end{array}
\right.
\end{eqnarray}
Therefore, $z=1$ is a well-defined limit of the lifshitz-Yang-Mills black holes. \\ 
We can conclude from the constraints \eqref{con1} that in order to support the Lifshitz spacetime, we need to fix the coupling constants of the two Yang-Mills gauge fields. It also shows that, for the hyperbolic and dS solutions with the conditions $k=-1$ and $\Lambda>0$, we terminate to the imaginary values for $e_{1}$ and $e_{2}$ in the case $z\neq 1$. So, in order to refuse this condition, we consider just the spherical and AdS solutions. By these conditions, the obtained solutions \eqref{ff} have the asymptotic behavior $f(r)\rightarrow 1$ as $r\rightarrow \infty$. \\
Now, we would like to investigate the physical properties of the Lifshitz-Yang-Mills-dilaton black hole solutions. If we calculate the Kretschmann scalar $R_{\alpha\beta\gamma\zeta}R^{\alpha\beta\gamma\zeta}$, it goes to infinity as $r\rightarrow 0$. Therefore, it announces of an essential singularity located at the origin for the mentioned black hole. We can also obtain the event horizon radius $r_{+}$ of the solutions by fixing $f(r_{+})=0$. This leads to  
\begin{eqnarray}\label{mas}
m(r_{+})&=& r_{+}^{z+n-1}+\frac{L^2(n-2)r_{+}^{z+n-3}}{z (z+n-3)}+\left\{
\begin{array}{ll}
$$\frac{(n-2)L^2 e_{3}^2 r_{+}^{-z+n-3}}{(z-n+3)}s^{2(z-1)}$$,\quad\quad\quad\quad\quad\quad \quad  \ {\mathrm{for}\,\, z\neq n-3 }\quad &  \\ \\
$$-(n-2)L^2 e_{3}^2r_{+}^{-z+n-3}\mathrm{ln}(r_{+})s^{2(z-1)}$$,\quad\quad  \ {\mathrm{for}\,\, z=n-3.}\quad &
\end{array}
\right.
\end{eqnarray}
According to the above relation and depending on the values of the parameters $e_{3}$, $m$, $z$, and $n$, the solutions can indicate a black hole with two horizons, an extreme black hole, or a naked singularity. To describe this behavior, we have plotted function $m(r_{+})$ versus $r_{+}$ for two fixed values of the mass parameter ($m=5.1,3.5$) and different values of $e_{3}$, $z$, and $n$ in Fig. \ref{Fig1}. In Fig.\ref{Fig1a}, there is an extreme black hole for the mentioned parameters with $m=5.1$ and $z=3$, while there is a black hole with two horizons $r_{-}$ and $r_{+}$ for $z>3$ and a naked singularity for $z<3$. In Fig. \ref{Fig1b} with $z=2$, an extreme five-dimensional black hole can be created for $m=5.1$, if we choose $e_{3}$ in the range $1<e_{3}<2$. So, a black hole with two horizons happens for small values of $e_{3}$. All three figures indicate that it may be more possible to see a black hole with inner and outer horizons if we choose a large value for the parameter $m$. \\
In the next section, we are going to study the thermodynamic characteristic of the Lifshitz-Yang-Mills-dilaton black hole.
\section{Thermodynamic behaviors of the Lifshitz-Yang-Mills-dilaton black hole}\label{thermo1}
Based on the AdS/CFT correspondence, one can use the thermodynamic properties of an AdS black hole to study its dual system behaviors in the conformal field theory. Some phenomena, such as the Nernst effect \cite{Hart2}, quantum Hall effect \cite{Hart3}, and superconductors \cite{Hart0,Hart1,Hart11,Hart33,Ammon,B1,B2,B3}, have been analyzed by this correspondence. In this section, we aim to evaluate the thermodynamic quantities such as mass, temperature, Yang-Mills charge, and the electric potential of the Lifshitz-Yang-Mills-dilaton black hole and probe the first law of thermodynamics. We also investigate the thermal stability of the obtained solutions in the grand canonical ensemble. It is hoped that the Lifshitz-Yang-Mills-dilaton black hole thermodynamics will be good candidates for better understanding condensed matter systems. \\
\subsection{Thermodynamic quantities and first law}
In order to obtain the mass of the Lifshitz-Yang-Mills-dilaton black hole, we use the modified subtraction method of Brown and York (BY) \cite{Brown,Brown2}. In this formalism, we should write the metric \eqref{metric} as below:
\begin{eqnarray}\label{metric2}
ds^2=-A(\mathcal{R}) dt^2+\frac{d\mathcal{R}^2}{B(\mathcal{R})}+\mathcal{R}^2 d\Omega_{n-1}^2,
\end{eqnarray}
where it requires $\mathcal{R}=r$ and  
\begin{eqnarray}\label{laps}
A(\mathcal{R})=\frac{r(\mathcal{R})^{2z}}{L^{2z}}f(r(\mathcal{R}))\,\,\,\,\,\,\, \mathrm{and}\,\,\, B(\mathcal{R})=\frac{r(\mathcal{R})^2}{L^2}f(r(\mathcal{R})).
\end{eqnarray}
It is a matter of calculation to show the quasilocal mass as
\begin{eqnarray}\label{mass1}
M_{T}=\frac{1}{8\pi}\int_{\Sigma} d^{n-1}x\sqrt{\sigma}\{(K_{ab}-K \gamma_{ab})\}n^{a}\xi^{b},
\end{eqnarray} 
where $K_{ab}$ is the extrinsic curvature of the metric and $\sigma$ illustrates the determinant of $\sigma_{ab}$(the metric of the boundary $\Sigma$). We specify $n^{a}$ and $\xi^{b}$ as the timelike unit normal vector to and a timelike killing vector field on the boundary $\Sigma$, respectively.
In the mass calculation process \eqref{mass1}, some divergences appear when we use the limit $r\rightarrow\infty$. To avoid this problem, we consider a background metric such as \eqref{metric2} with definitions \eqref{laps} and 
\begin{eqnarray}
f_{0}(r(\mathcal{R}))=1+\frac{L^2(n-2)}{z(z+n-3)[r(\mathcal{R})]^2}.
\end{eqnarray}
We obtain the mass of the background metric as below:
\begin{eqnarray}
M_{0}=\frac{1}{8\pi}\int_{\Sigma} d^{n-1}x\sqrt{\sigma}\{(K^{0}_{ab}-K^{0} \gamma^{0}_{ab})\}n^{a}\xi^{b},
\end{eqnarray} 
where $K^{0}_{ab}$ is the extrinsic curvature of the related metric.
Finally, the mass of the Lifshitz-Yang-Mills-dilaton black hole is obtained: 
\begin{eqnarray}\label{mm1}
M=M_{T}-M_{0}=\frac{(n-1)\omega_{n-1}}{16\pi L^{z+1}}m,
\end{eqnarray}
where $m$ is displayed in Eq. \eqref{mas} and $\omega_{n-1}$ is the volume of the $(n-1)$-dimensional hypersurface $\Sigma$. The temperature of this black hole at the outer horizon $r_{+}$ may be obtained as follows:
\begin{eqnarray}\label{Temp}
T_{+}&=&\frac{\kappa}{2\pi}=\frac{1}{2\pi}\sqrt{-\frac{1}{2}(\nabla_{\mu}\chi_{\nu})(\nabla_{\nu}\chi_{\mu})}\,|_{r=r_{+}}=\frac{r^{z+1}f^{'}(r)}{4\pi L^{z+1}}|_{r=r_{+}}\nonumber\\
&=&\frac{(n-2)}{4\pi z L^{z-1}}r_{+}^{z-2}+\frac{(z+n-1)}{4\pi L^{z+1}}r_{+}^{z}-\frac{(n-2) e_{3}^2 s^{2(z-1)}}{4\pi L^{z-1} r_{+}^{z+2}}.
\end{eqnarray}
It shows that, unlike the lap function $f(r)$, the temperature has just one form for the two cases $z\neq n-3$ and $z=n-3$. 
Using the so-called area law of entropy which is usable for nearly most black holes \cite{Beck,Haw,Haw2,Haw3}, we obtain the entropy of the Lifshitz-Yang-Mills-dilaton black hole as below: 
\begin{eqnarray}\label{entr1}
S=\frac{A}{4}=\frac{r_{+}^{n-1}}{4}\omega_{n-1}. 
\end{eqnarray}
We can also calculate the Yang-Mills charge of this black hole from the Gauss law:
\begin{eqnarray}\label{charg1}
Q=\frac{1}{4\pi\sqrt{(n-1)(n-2)}}\int d^{n-1}x\sqrt{Tr(F_{\mu\nu}^{(a)}F_{\mu\nu}^{(a)})}=\frac{\omega_{n-1}}{4\pi L^{1-z}}e_{3}.
\end{eqnarray}
In the purpose of probing the validity of the first law of thermodynamics, we consider the mass $M$ as a function of the extensive parameters $S$ and $Q$, where  
\begin{eqnarray}\label{T&U}
T=\bigg(\frac{\partial M}{\partial S}\bigg)_{Q},\,\,\,\,\, U=\bigg(\frac{\partial M}{\partial Q}\bigg)_{S}.
\end{eqnarray} 
The obtained results illustrate that the calculated value for $T$ in \eqref{T&U} is equal to the temperature Eq. \eqref{Temp}. So, the first law of thermodynamics is established
\begin{eqnarray}
dM=TdS+UdQ,
\end{eqnarray} 
where, using the relations \eqref{mas}, \eqref{mm1}, \eqref{charg1}, and \eqref{T&U}, the electric potential $U$ is obtained as follows:
\begin{eqnarray}
U=\bigg(\frac{\partial M}{\partial Q}\bigg)_{S}=-\frac{(n-1)(n-2)s^{2z-2}}{2 r_{+}^{z-n+3}L^{2z-2}}e_{3}\times\left\{
\begin{array}{ll}
$$\frac{1}{-z+n-3}$$,\quad\quad\quad\quad\quad\quad \quad  \ {\mathrm{for}\,\, z\neq n-3 }\quad &  \\ \\
$$\mathrm{ln}(r_{+})$$,\quad\quad\quad\quad \quad\quad\quad\quad\ {\mathrm{for}\,\, z=n-3.}\quad &
\end{array}
\right.
\end{eqnarray}
\subsection{Thermal stability}\label{thermal1}
\begin{figure}
\centering\subfigure[$det(H)$]{\includegraphics[scale=0.27]{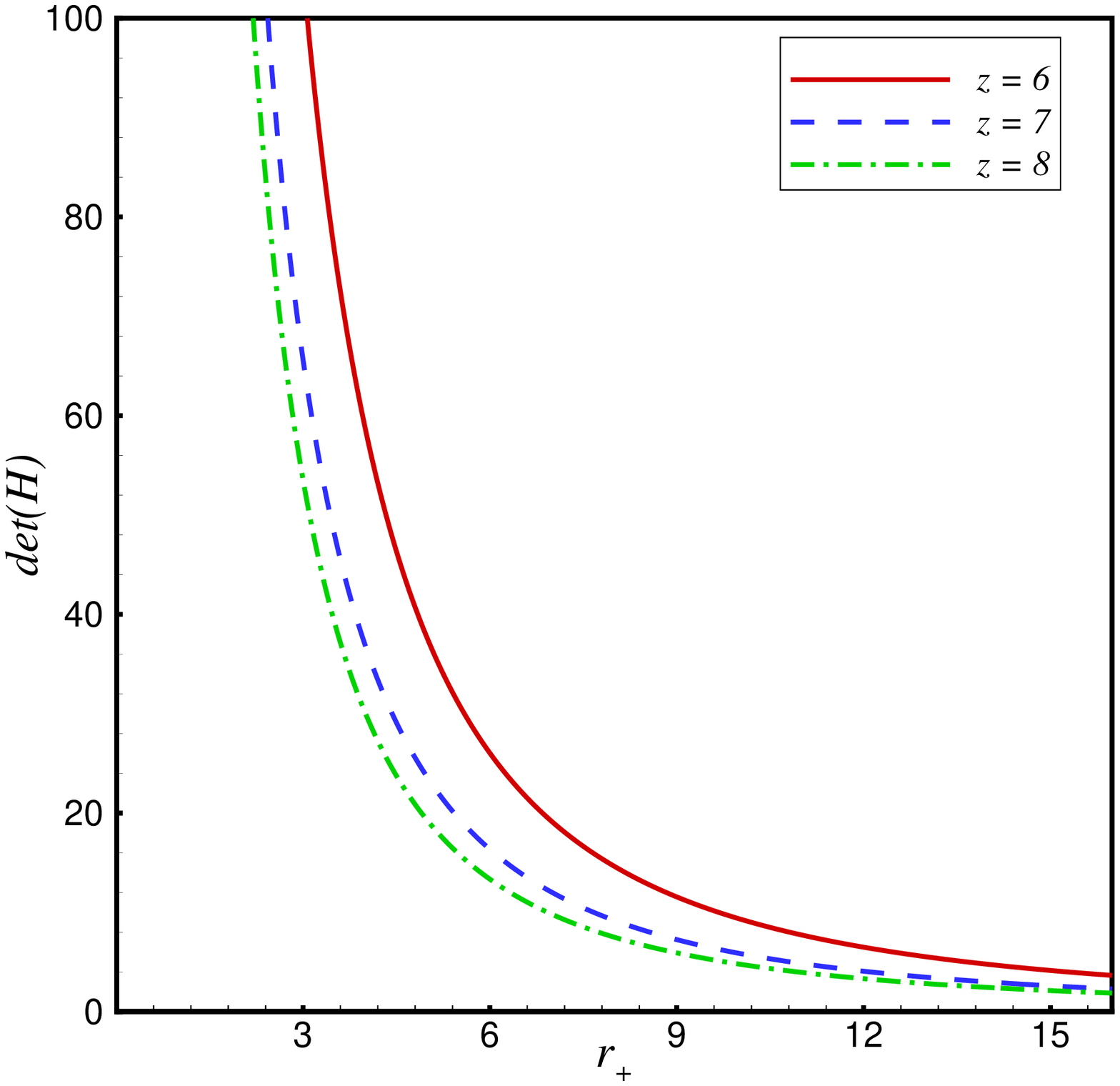}\label{Fig2a}}\hspace*{.2cm}
\subfigure[$d^2 M/dS^2$ ]{\includegraphics[scale=0.27]{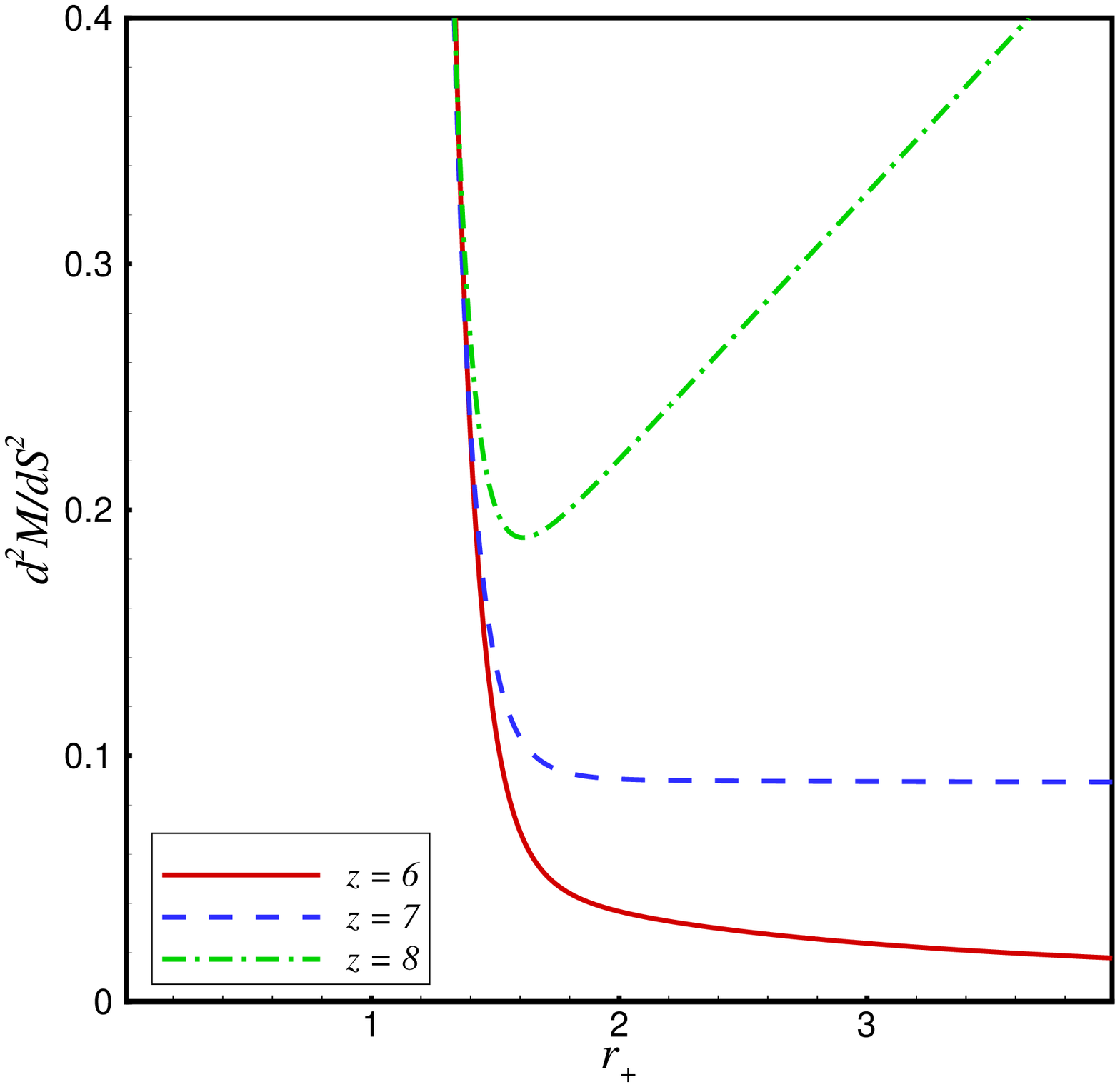}\label{Fig2b}}\hspace*{.2cm}
\subfigure[Temperature$T_{+}$]{\includegraphics[scale=0.27]{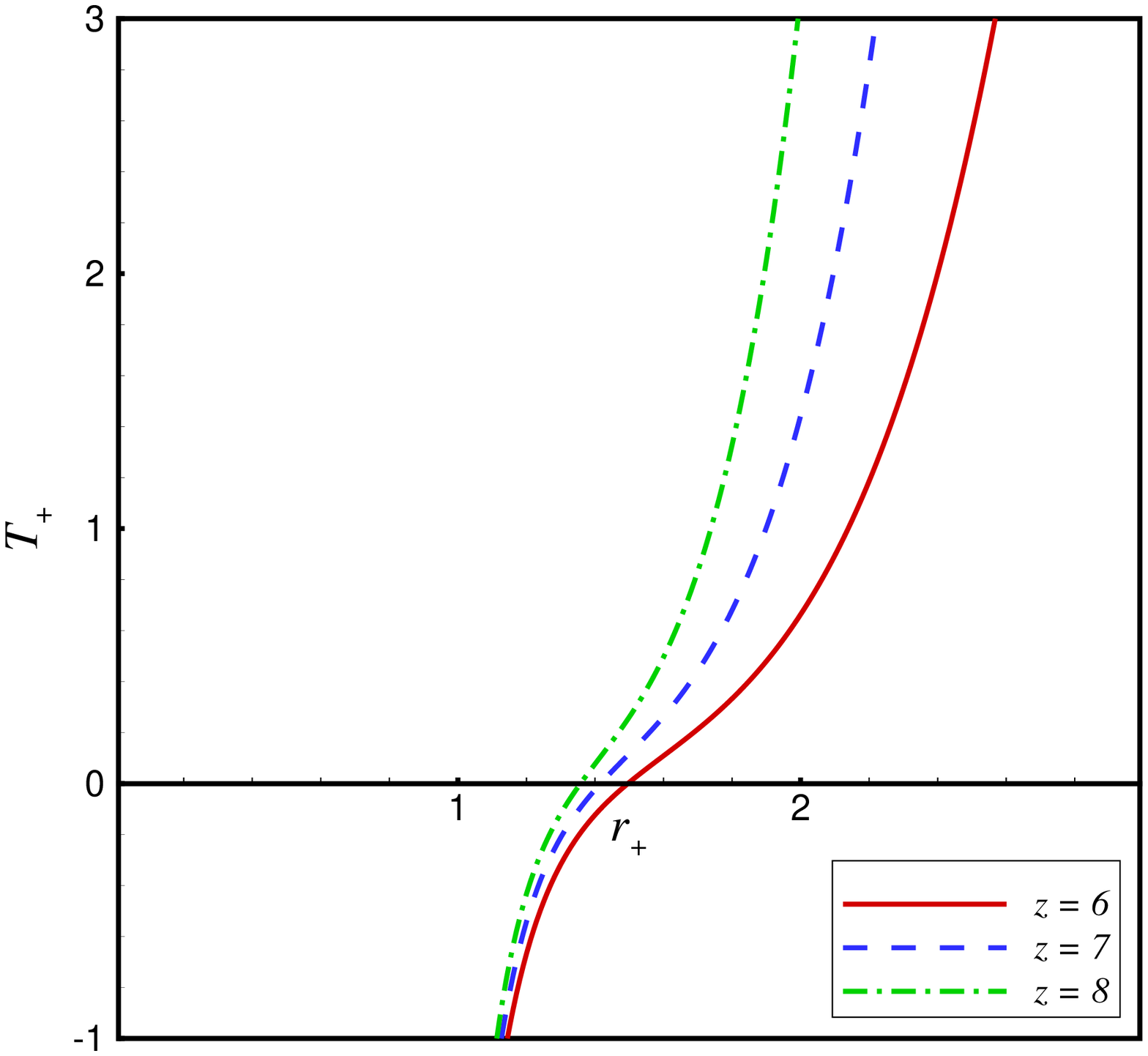}\label{Fig2c}}\caption{Thermal stability with respect to $r_{+}$ for different $z$ with $Q=2$ and $n=8$.}\label{Fig2}
\end{figure}

\begin{figure}
\centering
\subfigure[$det(H)$]{\includegraphics[scale=0.27]{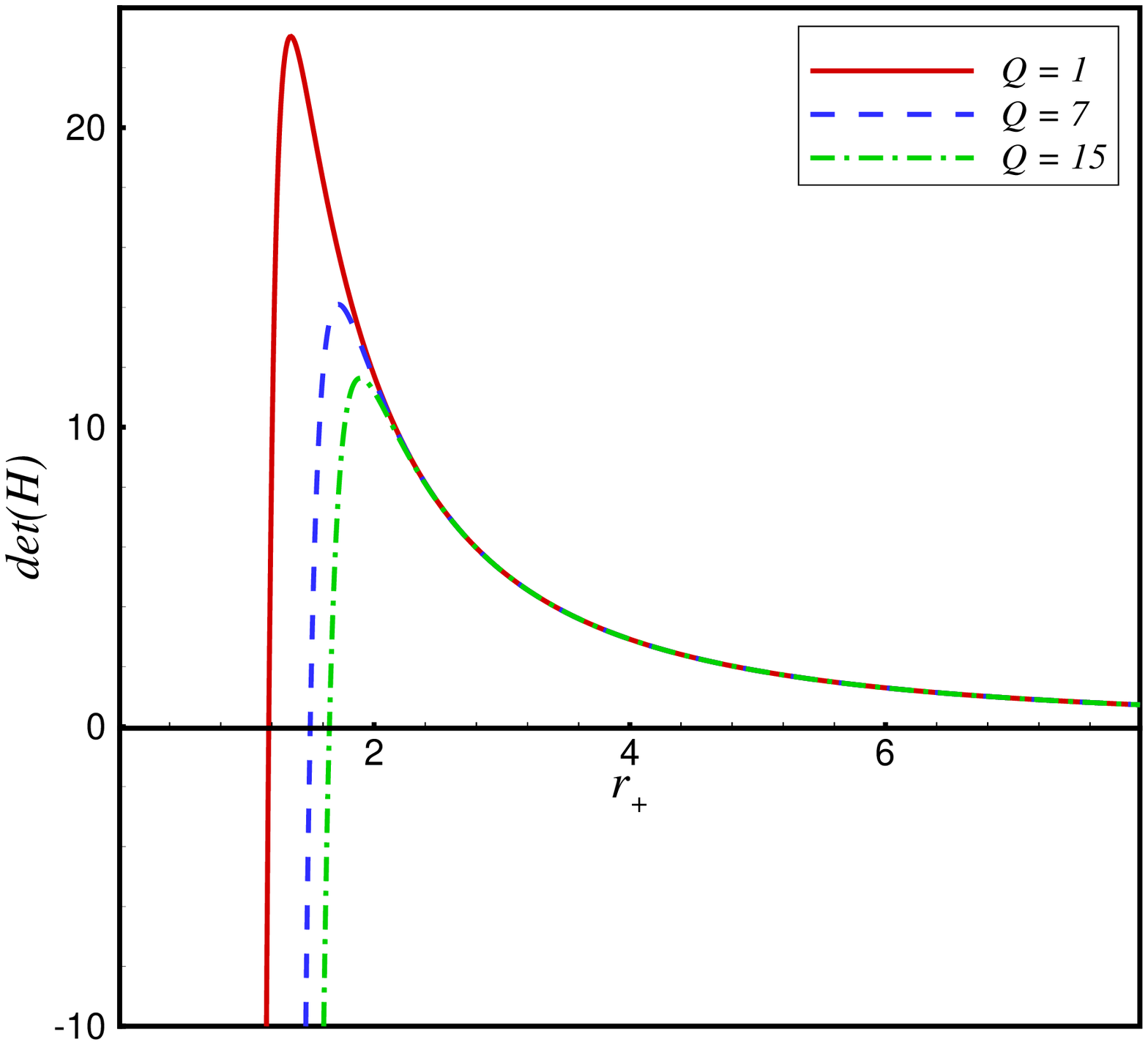}\label{Fig3a}}\hspace*{.2cm}
\subfigure[$d^2 M/dS^2$ ]{\includegraphics[scale=0.27]{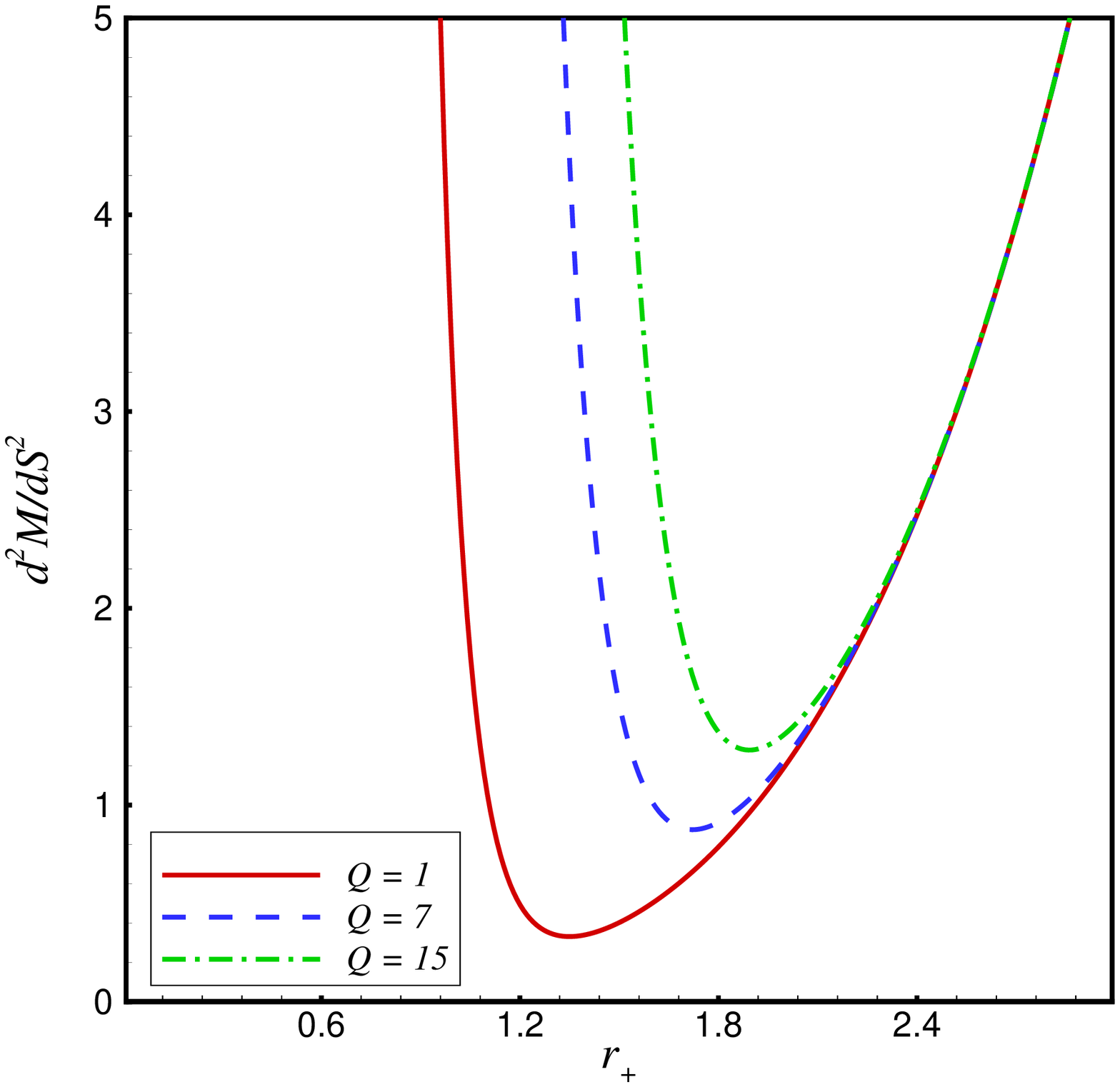}\label{Fig3b}}\hspace*{.2cm}
\subfigure[Temperature$T_{+}$]{\includegraphics[scale=0.27]{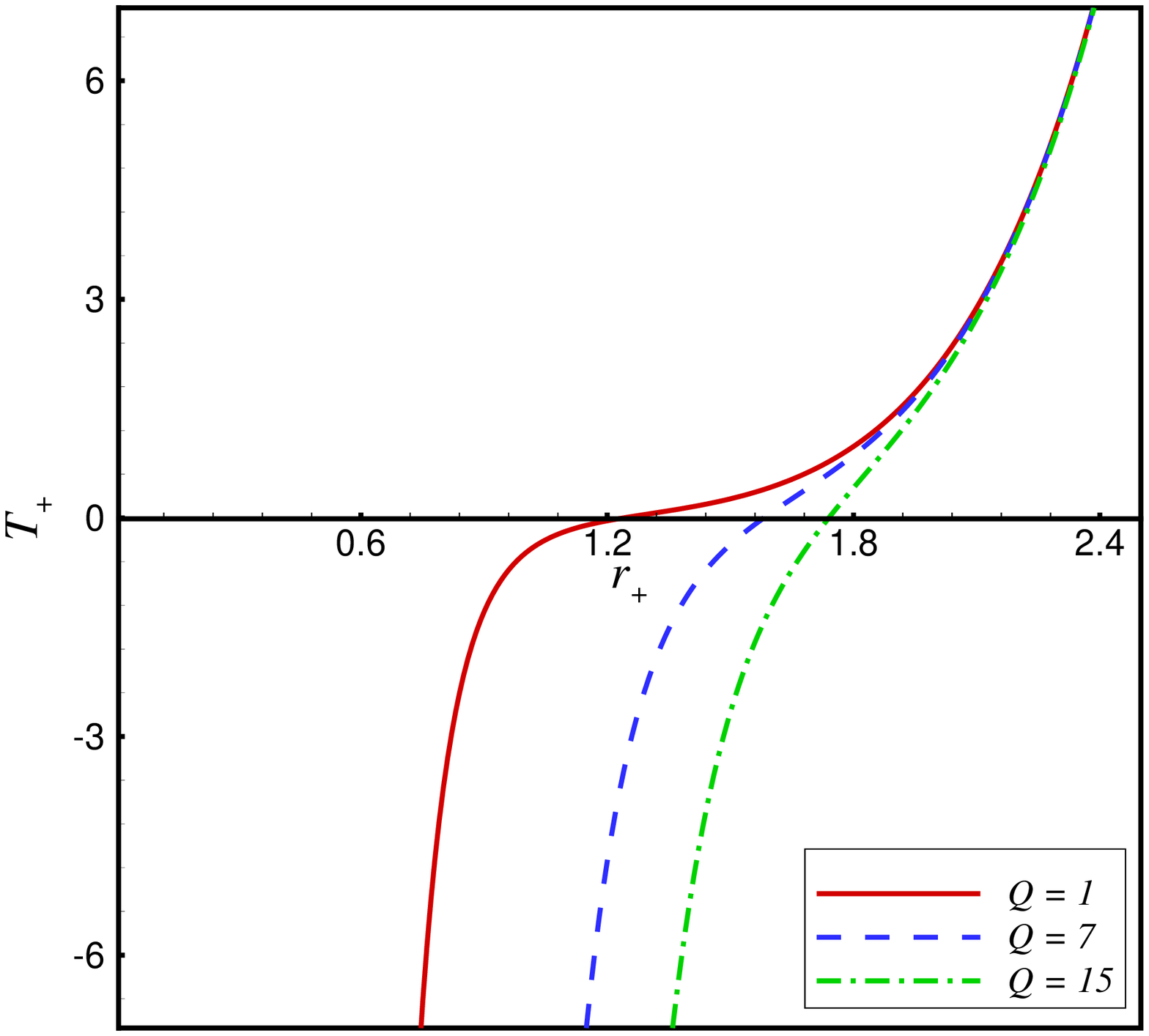}\label{Fig3c}}\caption{Thermal stability with respect to $r_{+}$ for different charge $Q$ with $z=7$ and $n=4$.}\label{Fig3}
\end{figure}

\begin{figure}
\centering
\subfigure[$det(H)$]{\includegraphics[scale=0.27]{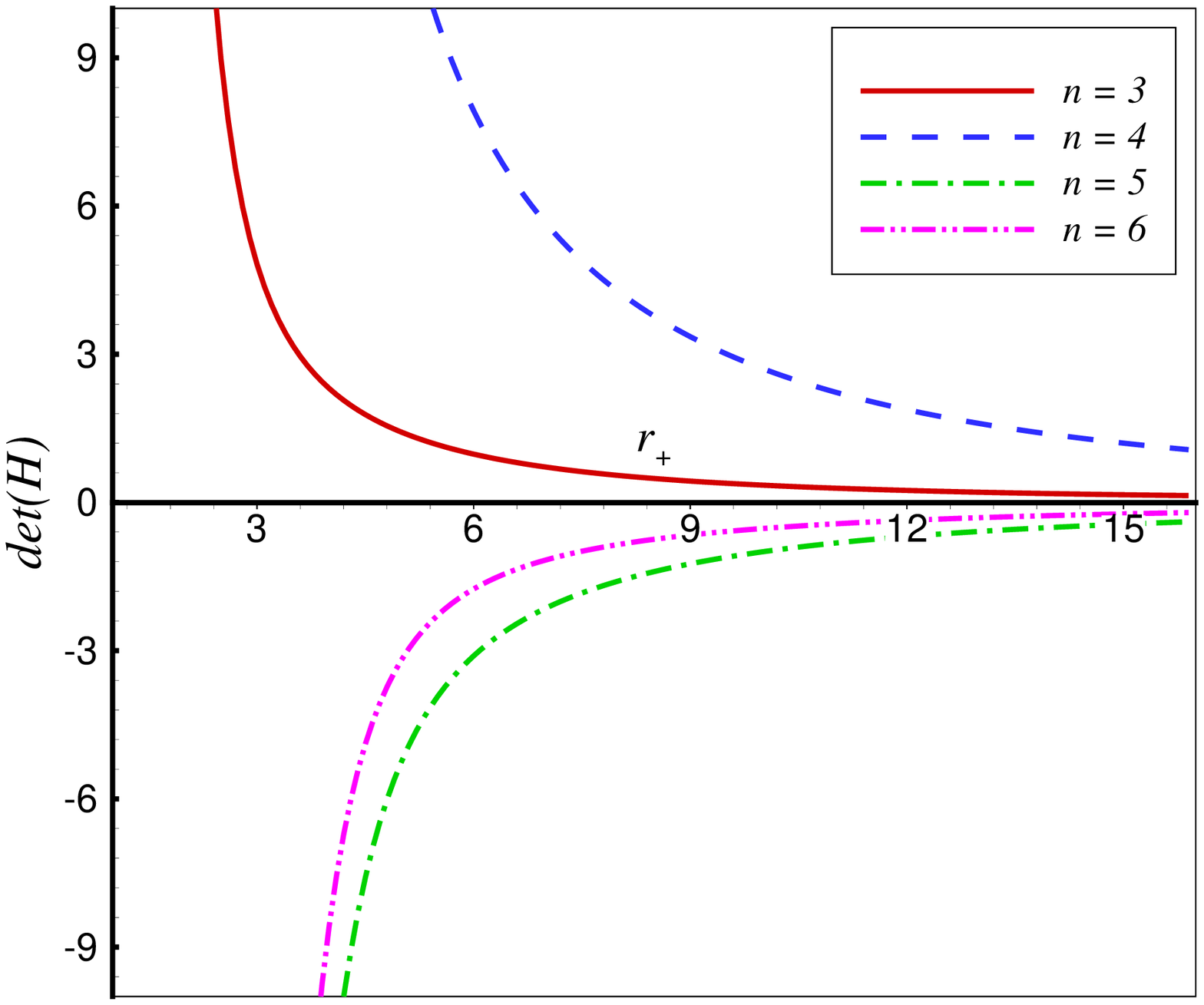}\label{Fig4a}}\hspace*{.2cm}
\subfigure[$d^2 M/dS^2$ ]{\includegraphics[scale=0.27]{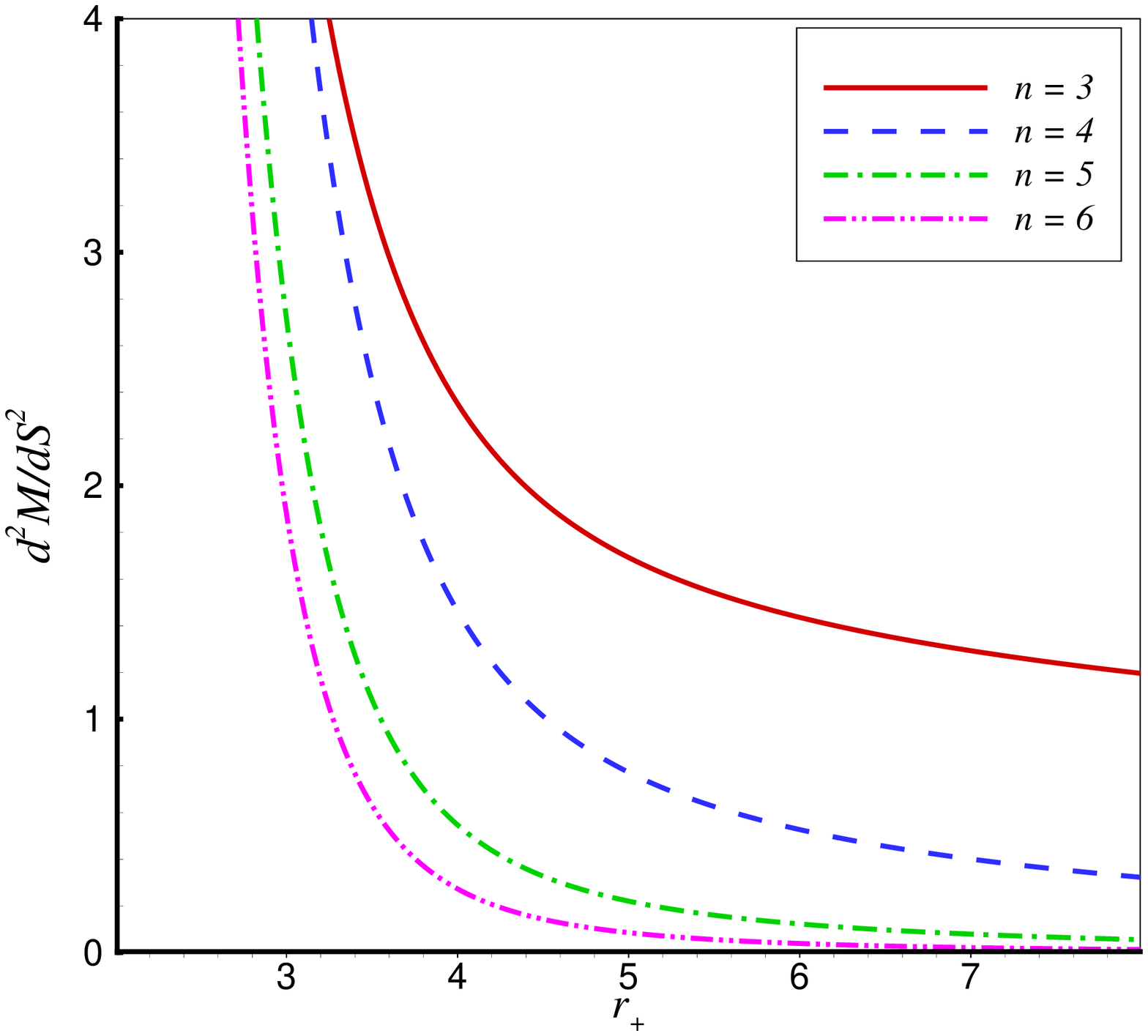}\label{Fig4b}}\hspace*{.2cm}
\subfigure[Temperature$T_{+}$]{\includegraphics[scale=0.27]{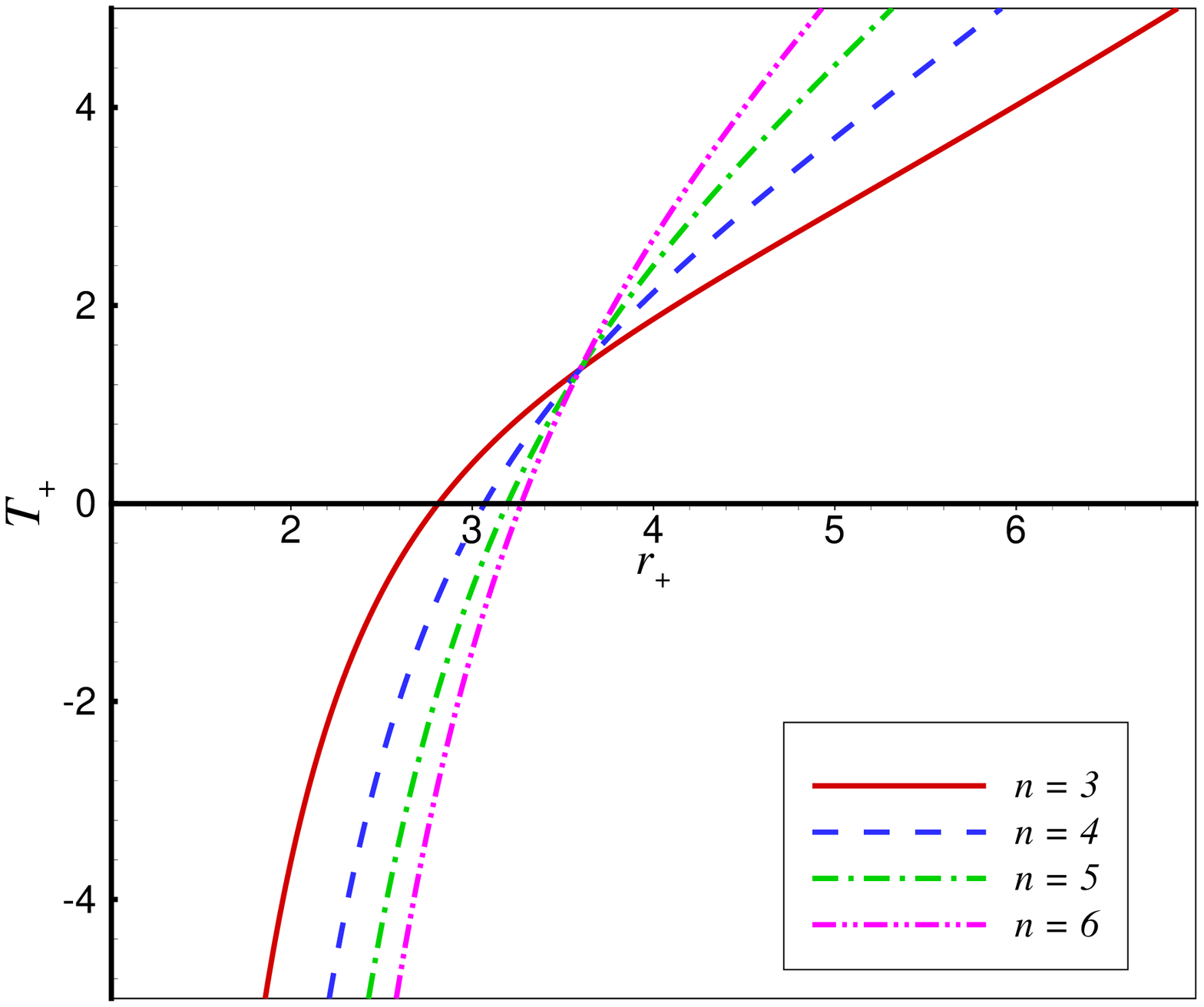}\label{Fig4c}}\caption{Thermal stability with respect to $r_{+}$ for different dimension $n$ with $z=1.5$ and $Q=2$.}\label{Fig4}
\end{figure}

Here, we want to study the thermal stability of the Lifshitz-Yang-Mills-dilaton black hole in the grand canonical ensemble for $z\geq1$. We probe the behavior of the energy $M(S,Q)$ with respect
to small variations of the entropy $S$ and charge $Q$. This study can specify whether the Lifshitz-Yang-Mills-dilaton black hole exists physically or not. In the grand canonical ensemble, the two parameters $S$ and $Q$ are variables, and so the positive value of the Hessian matrix determinant may lead to thermal stability. This matrix is found as
\begin{eqnarray}
H=\left[
\begin{array}{ccc}
\Big(\frac{\partial ^2 M}{\partial S^2}\Big)_{Q} & \Big(\frac{\partial ^2 M}{\partial S\partial Q}\Big)\\
\Big(\frac{\partial ^2 M}{\partial Q\partial S}\Big) & \Big(\frac{\partial ^2 M}{\partial Q^2}\Big)_{S}
\end{array} \right].
\end{eqnarray}
We obtain the Hessian matrix determinant of this black hole [we abbreviate it to det$H$] as follows: 
\begin{eqnarray}\label{detH1}
\det H&=&-\frac{2(n-2)^2s^{2z-2}}{(-z+n-3)L^{4z-2}}\bigg(\frac{(z-2)L^2}{zr_{+}^4}+\frac{z(z+n-1)}{(n-2)r_{+}^2}+\frac{16(2n-z-4)\pi^2 Q^2s^{2z-2}}{L^{2z-4}r_{+}^{2z+4}}\bigg), \,\,\,\,\,\,\, \mathrm{for}\,\, z\neq n-3
\end{eqnarray}
\begin{eqnarray}\label{detH2}
\det  H&=&-\frac{2(n-2)^2s^{2z-2}\mathrm{ln}(r_{+})}{L^{4z-2}}\bigg(\frac{(z-2)L^2}{zr_{+}^4}+\frac{z(z+n-1)}{(n-2)r_{+}^2}+\frac{16(2n-z-4)\pi^2 Q^2s^{2z-2}}{L^{2z-4}r_{+}^{2z+4}}\bigg)\nonumber\\
&&-\frac{64(n-2)^2\pi^2 Q^2s^{4z-4}}{L^{6z-6}r_{+}^{2z+4}},\,\,\,\,\,\,\,\,\,\,\,\,\,\,\,\,\,\,\,\,\,\,\,\,\,\,\,\,\,\,\,\,\,\,\,\,\,\,\,\,\,\,\,\,\,\,\,\,\,\,\,\,\,\,\,\,\,\,\,\,\,\,\,\,\,\,\,\,\,\,\,\,\,\,\,\,\,\,\,\,\,\,\,\,\,\,\,\,\,\,\,\,\,\,\,\,\,\,\,\,\,\,\,\,\,\,\,\,\,\,\,\,\,\,\,\,\,\,\,\,\,\,\,\,\,\,\,\,\,\,\,\,\,\,\,\,\,\,\,\,\,\,\,\,\,\,\,\,\,\,\,\,\, \mathrm{for}\,\, z= n-3.
\end{eqnarray}
The obtained result in Eq. \eqref{detH1} shows that, for $z<n-3$ and $z\geq2$, all three terms in the parentheses are non-negative, which leads to $\det H<0$. Therefore, the solutions with $z<n-3$ and $z\geq2$ have no stable regions. For $z>n-3$, the phrase before the parentheses is positive, and so the expressions in the parentheses determine the positive value of $\det(H)$. For large values of $r_{+}$, the contribution of the second term in the parentheses gets dominant. Since this term is positive, so the solutions with $z>n-3$ may be stable for large $r_{+}$. However, for small $r_{+}$, the third term has a dominant contribution. If $2n-z-4>0$, $\det H$ is positive, and if $2n-z-4<0$, $\det H$ is negative for small $r_{+}$. For the solutions with $z=n-3$ and $z\geq 2$ in Eq. \eqref{detH2}, $\det H$ is negative, and there is no thermal stability. \\
The other necessary conditions in order to have thermal stability in the grand canonical ensemble are the positive values of the quantities $\big(\partial ^2 M/\partial S^2\big)_{Q}$ and $T_{+}$. For a better review on the thermal stability of the Lifshitz-Yang-Mills-dilaton black hole, we have plotted $T_{+}$, $\big(\partial ^2 M/\partial S^2\big)_{Q}$, and $\det (H)$ in Figs.\ref{Fig2} -\ref{Fig4} for $L=s=1$. In Figs. \ref{Fig2} and \ref{Fig3}, we just consider the case $z>n-3$, which may lead to thermal stability. For different values of the exponent $z$ with $n=8$ in Fig. \ref{Fig2a}, $\det H$ is positive for all values of $r_{+}$. This is in accordance with our above statement. For these parameters, $\big(\partial ^2 M/\partial S^2\big)_{Q}$ is positive in Fig. \ref{Fig2b}, and so the sign of the temperature determines the thermal stability. In Fig. \ref{Fig2c}, there is a $r_{+\mathrm{min1}}$ where $T_{+}$ is positive for $r_{+}>r_{+\mathrm{min1}}$, and it decreases as the dynamical exponent $z$ increases. So the solutions with $z>n-3$ and $2n-z-4>0$ have a larger stable region if we choose a large value for $z$. \\
In Fig.\ref{Fig3}, we probe the thermal stability of the solutions with different charges $Q$ for the conditions $z>n-3$ and $2n-z-4<0$. In this figure, $\big(\partial ^2 M/\partial S^2\big)_{Q}$ is positive for all values of $r_{+}$, and so we should take a unit positive region between $T_{+}$ and $\det (H)$. The results represent that the black hole is stable for large $r_{+}$, and this stable region increases if we choose small charge values. \\
Although the exponent $z$ is an integer, it may be interesting to speak about the solutions with $1\leq z<2$, which are in the category $z\neq n-3$. So, we have probed the thermal stability of the solutions with $z=1.5$ in Fig.\ref{Fig4}. This figure shows that, for the solutions with $z<2$, $\det (H)$ is positive only for dimensions $n=3,4$. For these values, there is a $r_{+\mathrm{min2}}$ which the temperature is positive for $r_{+}>r_{+\mathrm{min2}}$, and so thermal stability happens. 

\section{Critical behavior of the Lifshitz-Yang-Mills-dilaton black hole}\label{PV1}
\begin{figure}
\centering
\subfigure[$z=1.2$, $n=3$ ]{\includegraphics[scale=0.27]{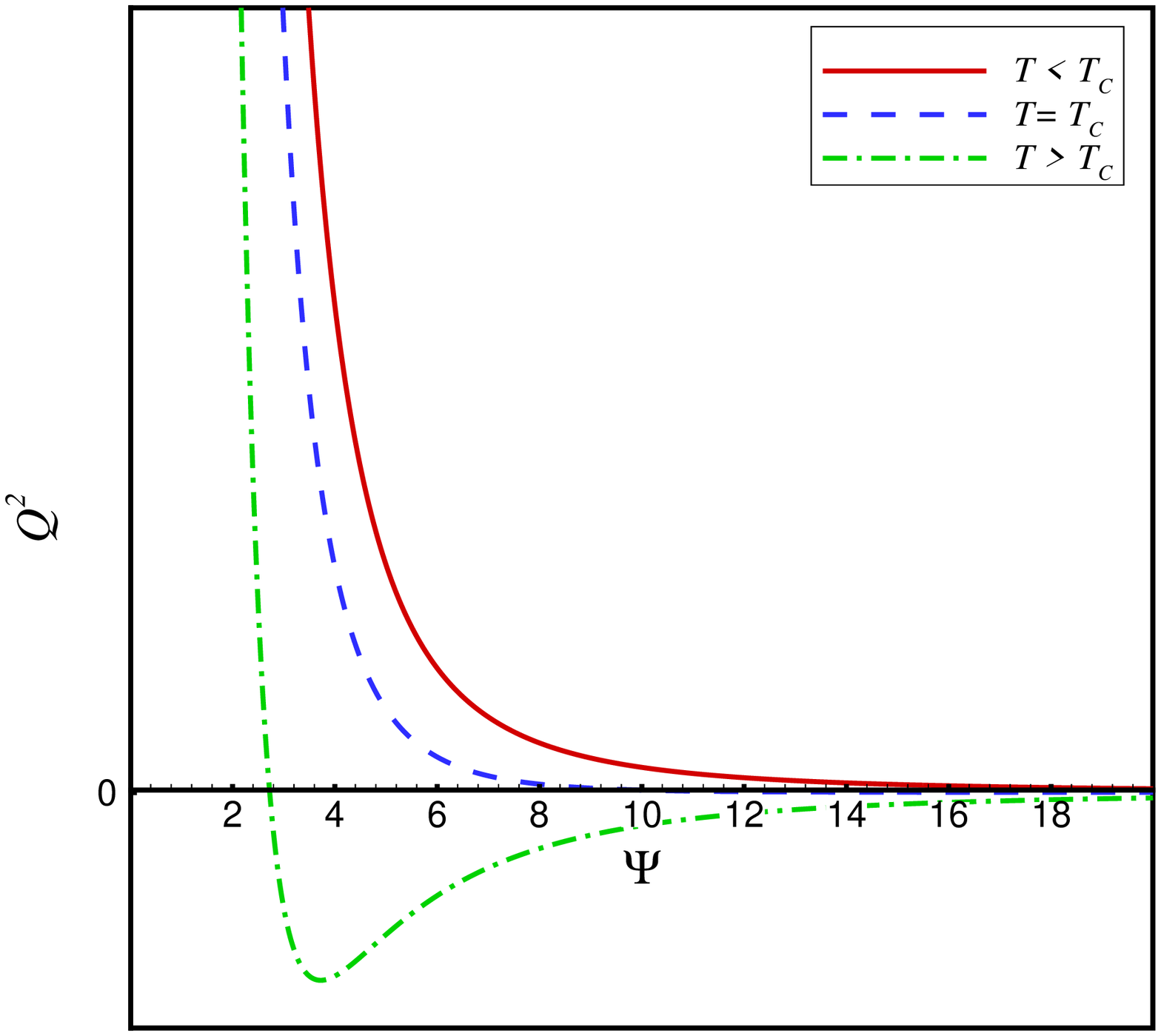}\label{Fig5a}}\hspace*{.2cm}
\subfigure[$z=1.7$, $n=4$]{\includegraphics[scale=0.27]{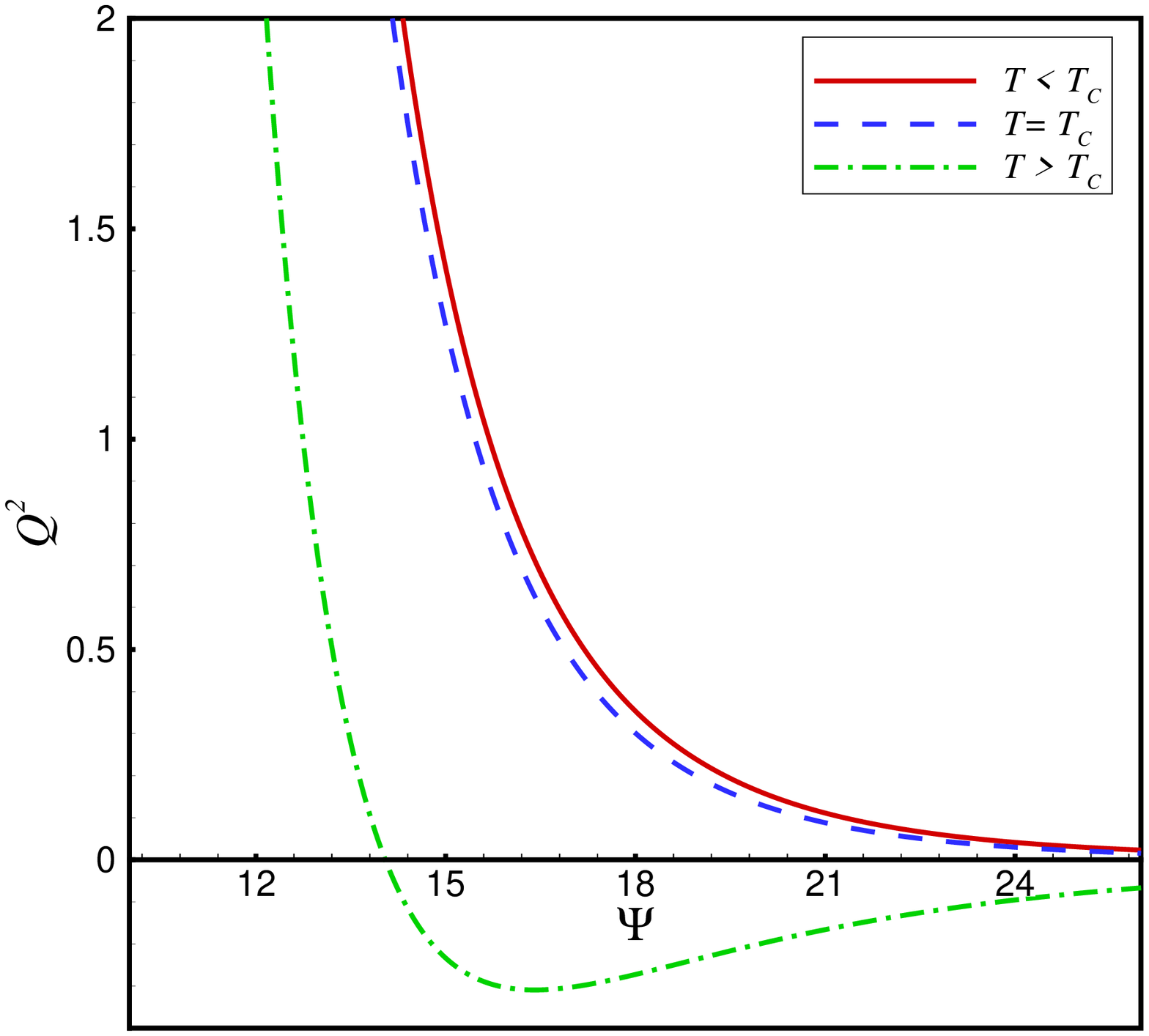}\label{Fig5b}}\caption{The isothermal $Q^2-\Psi$ diagram for $s=1$ and $b=1$.}\label{Fig5}
\end{figure}

\begin{figure}
\centering
\subfigure[$z=1.1$, $n=3$ ]{\includegraphics[scale=0.27]{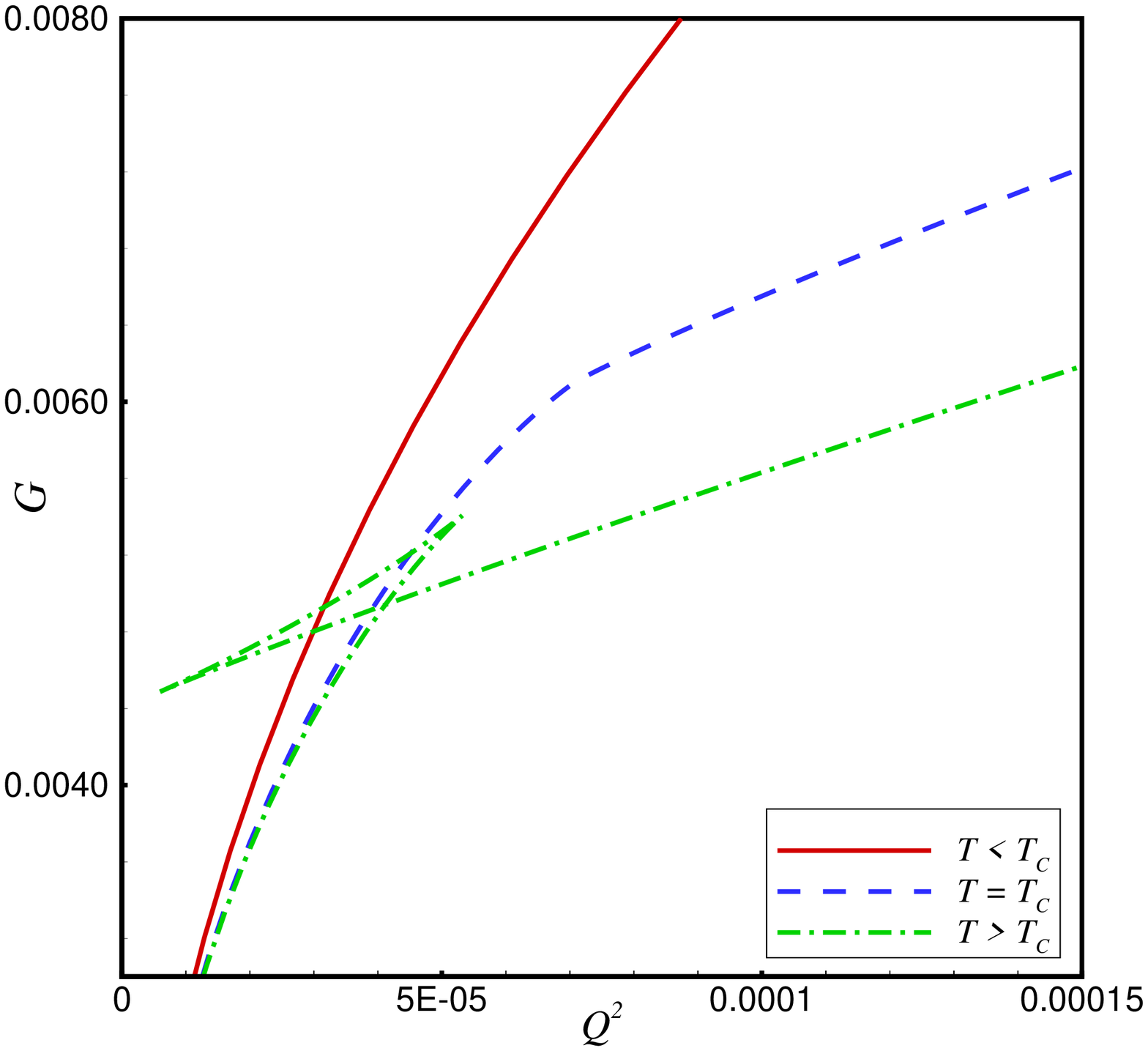}\label{Fig6a}}\hspace*{.2cm}
\subfigure[$z=1.6$, $n=4$]{\includegraphics[scale=0.27]{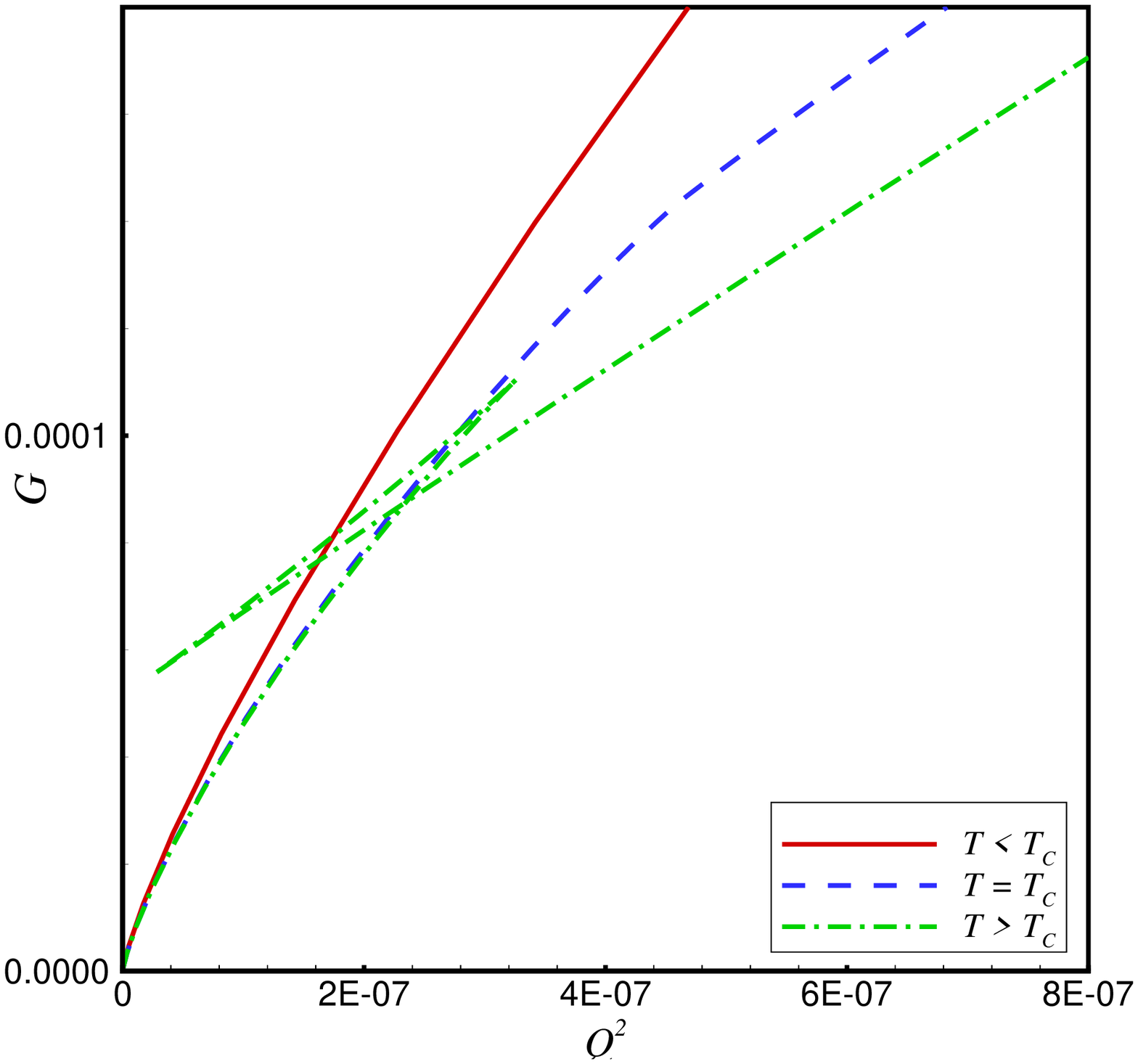}\label{Fig6b}}\caption{Gibbs free energy versus $Q^2$ for $s=2$ and $b=1$.}\label{Fig6}
\end{figure}

Critical behavior is another interesting thermodynamic issue which can be studied for a black hole. The first step in this field was done by Hawking and Page that showed a certain phase transition for the Schwarzschild-AdS black hole \cite{Haw1}. Then, the critical behavior of the Reissner-Nordstr$\rm\ddot{o}$m-AdS black hole was studied in Refs. \cite{Cham,Liu}. The authors of these papers showed a first-order phase transition similar to the van der Waals liquid gas one. In Refs. \cite{Ura,Dolan,Cv,Mirza1,Mirza2}, the cosmological constant $\Lambda$ is considered as a variable thermodynamic pressure, and so an analytical equation of state such as $P=P(v,T)$ may be obtained. A glance at Refs. \cite{lif1,Shey12} reveals that it is not possible to have a Smarr relation for the charged Lifshitz black holes. This convinced the authors in Refs. \cite{Deh11,Deh22} to consider both the cosmological constant $\Lambda$ and the square charge $Q^2$ as thermodynamic variable parameters. It should be noted that mass should be defined as the enthalpy of the black hole in the extended phase space \cite{Ura,Kub,Dol1}. With these considerations, the first law of thermodynamics in the extended phase space is followed from 
\begin{eqnarray}\label{first}
dM=TdS+VdP+\Psi dQ^2,
\end{eqnarray}
where $\Psi$ is referred to the conjugate of $Q^2$ with definition  
\begin{eqnarray}\label{psi1}
\Psi=\bigg(\frac{\partial M}{\partial Q^2}\bigg)_{S,P}=-\frac{(n-1)(n-2)\pi s^{2z-2}}{(-z+n-3)L^{3z-3}}r_{+}^{-z+n-3}
\end{eqnarray}
and $M$ and $Q$ are defined in Eqs. \eqref{mm1} and \eqref{charg1}. So, the equation of state and the Smarr relation are obtained as $Q^2=Q^2(T,\Psi)$ and $M=M(S,Q^2,P)$, respectively. So,
if we consider the volume $V=r_{+}^n/n$, then its conjugate, pressure, is obtained from Eq. \eqref{mm1} as 
\begin{eqnarray}\label{PPP}
P=\frac{n(n-1)r_{+}^{z-1}}{16\pi L^{z+1}}.
\end{eqnarray}
By the above definitions and Eqs. \eqref{mm1} -\eqref{charg1}, \eqref{psi1}, and \eqref{PPP}, we can find a Smarr-type relation as below: 
\begin{eqnarray}
(z+n-3)M=(n-1)TS-2PV+2z\,\Psi Q^2,\,\,\,\,\,\,\,\,\,\,\,\,\,\,\,\,\,\,\,\mathrm{for}\, z\neq n-3,
\end{eqnarray}
which there is no Smarr relation for the case $z=n-3$. 
If we substitute Eqs. \eqref{charg1} and \eqref{psi1} in Eq. \eqref{Temp} and then solve the related equation, we can obtain the equation of state  
\begin{eqnarray}\label{QQ1}
Q^2&=&-\frac{L^{3z-3}T}{4\pi(n-2)s^{2z-2}}\bigg(\frac{(z-n+3)L^{3z-3}\Psi}{(n-1)(n-2)\pi s^{2z-2}}\bigg)^{\frac{z+2}{-z+n-3}}+\frac{L^{2z-2}}{16\pi^2 z s^{2z-2}}\bigg(\frac{(z-n+3)L^{3z-3}\Psi}{(n-1)(n-2)\pi s^{2z-2}}\bigg)^{\frac{2z}{-z+n-3}}\nonumber\\
&&+\frac{(z+n-1)L^{2z-4}}{16\pi^2(n-2)s^{2z-2}}\bigg(\frac{(z-n+3)L^{3z-3}\Psi}{(n-1)(n-2)\pi s^{2z-2}}\bigg)^{\frac{2(z+1)}{-z+n-3}}.
\end{eqnarray} 
In order to obtain the critical points, we must apply the conditions 
\begin{eqnarray}\label{QQ}
\frac{\partial Q^2}{\partial \Psi}\mid_{T_{C}}=0\,\,\,\,\,\,\,,\,\,\,\frac{\partial^2 Q^2}{\partial \Psi^2}\mid_{T_{C}}=0,
\end{eqnarray}
where they lead to $\Psi_{C}$ and the critical temperature $T_{C}$:  
\begin{eqnarray}
\Psi_{C}=-\frac{(n-1)(n-2)\pi s^{2z-2}}{(-z+n-3)L^{3z-3}}\bigg(-\frac{(z-2)(n-2)L^2}{z(z+1)(z+n-1)}\bigg)^{\frac{-z+n-3}{2}},
\end{eqnarray}
\begin{eqnarray}\label{Tcritical}
T_{C}=-\frac{(z+1)(z+n-1)}{(z-2)(z+2)\pi L^{z+1}}\bigg(-\frac{(z-2)(n-2)L^2}{z(z+1)(z+n-1)}\bigg)^{\frac{z}{2}}.
\end{eqnarray}
In order to have a real positive critical temperature, we should choose either $1\le z<2$ or an odd number value for $z/2$. As we explained in Sec. \ref{Field}, there is no ambiguity at $z=1$, and so the thermodynamic relationships are correct for this value. Our results show that the solutions with the condition $z/2$ do not manifest any transitions and so we consider only the first constraint $1\le z<2$. We have plotted the isotherms $Q^2-\Psi$ for the Lifshitz-Yang-Mills-dilaton black hole with the condition $1\le z<2$ in Fig. \ref{Fig5}. These figures demonstrate that there is a similarity between the isotherms of the Lifshitz-Yang-Mills-dilaton black hole and the ones in van der Waals liquid gas. \\
We can also probe the Gibbs free energy $G$ of the obtained black hole. It may be obtained by  
\begin{eqnarray}
G(Q^2,T)&=& M-TS
=\frac{(z+2)T}{4(-z+n-3)}r_{+}^{n-1}\nonumber\\
&&-\frac{(n-1)(n-2)}{8\pi(z+n-3)(-z+n-3)L^{z-1}}r_{+}^{z+n-3}-\frac{(n-1)(z+1)}{8\pi(-z+n-3)L^{z+1}}r_{+}^{z+n-1},\,\,\,\mathrm{for}\, z\neq n-3,
\end{eqnarray}
where, according to Eqs. \eqref{psi1} and \eqref{QQ1}, $r_{+}$ is a function of the temperature and the square charge, $r_{+}=r_{+}(T,Q^2)$. We have also plotted $G-Q^2$ diagrams for different values of $z$ and $n$ in Fig. \ref{Fig6}. The swallowtail behavior of the diagrams indicates that there is a first-order phase transition for $T>T_{C}$.

\section{The main structure of the Yang-Mills-dilaton black holes with a hyperscaling violation}\label{field2}
In this section, we would like to extend the Lifshitz solutions and obtain the hyperscaling violated Yang-Mills-dilaton black holes. As we said in Sec.\ref{intro}, not only do these solutions have a dynamical violation, but they also have a hyperscaling violation, too. We rewrite this geometry as below:  
\begin{eqnarray}
ds^2=r^{2\alpha}\bigg(-\frac{f(r)r^{2z}}{L^{2z}}dt^2+\frac{L^2 dr^2}{r^2f(r)}+r^2 d\Omega_{k}^2\bigg),
\end{eqnarray}
where $\alpha=-\theta/(n-1)$ is the hyperscaling violation parameter. Following from the previous sections for the Lifshitz solutions, we just consider the spherical hypersurface $k=1$ to gain real solutions. If we use the field equations \eqref{eq1} -\eqref{eq3} and consider a potential with a Liouville form for the dilaton field
\begin{eqnarray}
V(\phi)=2\Lambda e^{\lambda\phi},
\end{eqnarray}
then we get to a set of asymptotically Yang-Mills-dilaton black hole solutions with a hyperscaling violation as follows 
\begin{eqnarray}\label{phi2}
\Phi(r)=\frac{n-1}{2}\sqrt{(\alpha+1)(\alpha+z-1)}\mathrm{ln}\bigg(\frac{r}{b}\bigg),
\end{eqnarray}
\begin{eqnarray}\label{fff}
f(r)&=& 1+\frac{L^2(n-2)}{(z+2\alpha)(z+n-3+(n-1)\alpha)r^2}-\frac{m}{r^{z+(n-1)(\alpha+1)}}\nonumber\\
&&+\left\{
\begin{array}{ll}
$$\frac{(n-2)L^2 e_{3}^2}{(\alpha+1)(z-n+3-(n-5)\alpha)r^{2z+2+4\alpha}}b^{2(\alpha+z-1)}$$,\quad\quad\quad \quad  \ {\mathrm{for}\,\, z\neq n-3+(n-5)\alpha }\quad &  \\ \\
$$-\frac{(n-2)L^2 e_{3}^2}{(\alpha+1)r^{2z+2+4\alpha}}\mathrm{ln}(\frac{r}{r_{0}})b^{2(\alpha+z-1)}$$,\quad\quad\quad\quad  \ \,\,\,\,\,\,\,\,\,\,\,\,\,\,\,\,\,\,\,{\mathrm{for}\,\, z=n-3+(n-5)\alpha.}\quad &
\end{array}
\right.
\end{eqnarray}\\
where $b$ is a constant of integration and we choose $r_{0}=1$. For the above solutions, one may find
\begin{eqnarray}\label{con2}
\xi_{1}&=&-\frac{(n-1)\lambda\sqrt{(\alpha+1)(\alpha+z-1)}+8(\alpha+1)}{4\sqrt{(\alpha+1)(\alpha+z-1)}}\,\,\,\,\,,\,\,\,\,\xi_{2}=-\sqrt{\frac{\alpha+1}{\alpha+z-1}}\,\,\,,\,\,\,\xi_{3}=\sqrt{\frac{\alpha+z-1}{\alpha+1}},\nonumber\\
e_{1}^2&=&-\frac{2(z-1)\Lambda}{(n-1)(n-2)(2\alpha+z+1)}b^{4(\alpha+1)}\,\,\,,\,\,\,
e_{2}^2=\frac{\alpha+z-1}{z+2\alpha}b^{2(\alpha+1)},\nonumber\\
\Lambda&=&-\frac{(n-1)(2\alpha+z+1)(z+(n-1)(\alpha+1)}{4L^2}b^{-2\alpha}\,\,,\,\,\nonumber\\
\lambda&=&-\frac{4\alpha}{(n-1)\sqrt{(\alpha+1)(\alpha+z-1)}}.
\end{eqnarray}
For the limits $\lambda,\alpha\rightarrow 0$ and $b\rightarrow s$, the obtained solutions reduce to the Lifshitz ones in Eqs. \eqref{phi1} -\eqref{con1}. In order to have the asymptotic behavior $f(r)\rightarrow 1$ for the limit $r\rightarrow \infty$, the two conditions $z+(n-1)(\alpha+1)>0$ and $2z+2+4\alpha>0$ should be satisfied. Also, according to the relations in Eq. \eqref{con2}, the real solutions are accessible only for $(\alpha+1)(\alpha+z-1)>0$. They also show that we are not allowed to consider the cases $\alpha=-1$ or $\alpha=1-z$, since the solutions diverge. By these statements, if we consider the region $\alpha>0$, all constraints are satisfied. \\
If we return the transformation $\alpha=-\theta/(n-1)$, the hyperscaling Yang-Mills-dilaton black hole solutions are rewritten as below:
\begin{eqnarray}\label{phi3}
\Phi(r)=\frac{1}{2}\sqrt{(n-\theta-1)[(z-1)(n-1)-\theta]}\mathrm{ln}\bigg(\frac{r}{b}\bigg),
\end{eqnarray}
\begin{eqnarray}\label{ffff}
f(r)&=& 1+\frac{L^2(n-1)(n-2)}{[(n-1)z-2\theta](z+n-3-\theta)r^2}-\frac{m}{r^{z+n-\theta-1}}\nonumber\\
&&+\left\{
\begin{array}{ll}
$$\frac{(n-1)^2(n-2)L^2 e_{3}^2}{(\theta-n+1)[(n-z-3)(n-1)-(n-5)\theta]r^{2(z+1)-\frac{4\theta}{n-1}}}b^{2(z-1)-\frac{2\theta}{n-1}}$$,\quad\quad\quad \quad  \ {\mathrm{for}\,\, z\neq n-3-\frac{n-5}{n-1}\theta }\quad &  \\ \\
$$\frac{(n-1)(n-2)L^2 e_{3}^2}{(\theta-n+1)r^{2(z+1)-\frac{4\theta}{n-1}}}b^{2(z-1)-\frac{2\theta}{n-1}}\mathrm{ln}(r)$$,\quad\quad\quad\quad  \ \,\,\,\,\,\,\,\,\,\,\,\,\,\,\,\,\,\,\,\,\,\,\,\,\,\,\,\,\,\,\,\,\,\,\,\,\,\,\,\,\,\,\,\,\,{\mathrm{for}\,\, z=n-3-\frac{n-5}{n-1}\theta,}\quad &
\end{array}
\right.
\end{eqnarray}
where $\theta$ should be negative: $\theta<0$. For the above solutions, we obtain
\begin{eqnarray}\label{mas2}
m(r_{+})&=& r_{+}^{z+n-\theta-1}+\frac{L^2(n-1)(n-2)}{[(n-1)z-2\theta](z+n-3-\theta)}r_{+}^{z+n-\theta-3}\nonumber\\
&&+\left\{
\begin{array}{ll}
$$\frac{(n-1)^2(n-2)L^2 e_{3}^2}{(\theta-n+1)[(n-z-3)(n-1)-(n-5)\theta]}r_{+}^{n-z-\theta-3+\frac{4\theta}{n-1}}b^{2(z-1)-\frac{2\theta}{n-1}}$$,\quad\quad\quad \quad  \ {\mathrm{for}\,\, z\neq n-3-\frac{n-5}{n-1}\theta }\quad &  \\ \\
$$\frac{(n-1)(n-2)L^2 e_{3}^2}{(\theta-n+1)}r_{+}^{n-z-\theta-3+\frac{4\theta}{n-1}}b^{2(z-1)-\frac{2\theta}{n-1}}\mathrm{ln}(r_{+})$$,\quad\quad\quad\quad  \ \,\,\,\,\,\,\,\,\,\,\,\,\,\,\,\,\,\,\,{\mathrm{for}\,\, z=n-3-\frac{n-5}{n-1}\theta}\quad &
\end{array}
\right.
\end{eqnarray}
where, depending on the values of the parameters $m$, $n$, $z$, $e_{3}$, and $\theta$, the solutions may lead to a black hole with two horizons, an extreme black hole, or a naked singularity. \\
\section{Thermodynamic behaviors of the hyperscaling violated Yang-Mills-dilaton black hole}\label{thermo2}
Now, we want to study the thermodynamic properties of the hyperscaling violated Yang-Mills-dilaton black hole. Using the Brown and York subtraction formalism, the mass of this black hole is obtained as 
\begin{eqnarray}
M=\frac{(n-1-\theta)}{16\pi L^{z+1}}m,
\end{eqnarray}
where $m$ is defined in Eq. \eqref{mas2}.
The Hawking temperature and the entropy of the hyperscaling violated Yang-Mills-dilaton black hole are calculated as below:
\begin{eqnarray}\label{temp2}
T_{+}=\frac{r^{z+1}f^{'}(r)}{4\pi L^{z+1}}|_{r=r_{+}}=\frac{(z+n-\theta-1)r_{+}^z}{4\pi L^{z+1}}+\frac{(n-1)(n-2)r_{+}^{z-2}}{4\pi [z(n-1)-2\theta]L^{z-1}}+\frac{(n-1)(n-2) e_{3}^2 b^{2(z-1)-\frac{2\theta}{n-1}}}{4\pi(\theta-n+1) L^{z-1} r_{+}^{z+2-\frac{4\theta}{n-1}}},
\end{eqnarray}
\begin{eqnarray}
S=\frac{r_{+}^{n-1-\theta}}{4}.
\end{eqnarray}
Using the Gauss law in Eq. \eqref{charg1}, the charge is obtained as the following:
\begin{eqnarray}
Q=\frac{\omega_{n-1}}{4\pi L^{1-z}}e_{3}.
\end{eqnarray}
It is obvious that, for $\theta\rightarrow 0$, the obtained thermodynamic quantities are reduced to Eqs. \eqref{mm1} -\eqref{charg1}.
For the obtained solutions, the first law of thermodynamics is established, and the electric potential is followed from
\begin{eqnarray}
U=\bigg(\frac{\partial M}{\partial Q}\bigg)_{S}=-\frac{(n-1)(n-2)b^{2z-2-\frac{2\theta}{n-1}}}{2 r_{+}^{z-n+3+\frac{n-5}{n-1}\theta}L^{2z-2}}e_{3}\times\left\{
\begin{array}{ll}
$$\frac{n-1}{(n-1)(n-z-3)-\theta(n-5)}$$,\quad\quad\quad\quad\quad\quad \quad  \ {\mathrm{for}\,\, z\neq n-3 }\quad &  \\ \\
$$\mathrm{ln}(r_{+})$$,\quad\quad\quad\quad \quad\quad\quad\quad\quad\quad\quad\quad\quad\ {\mathrm{for}\,\, z=n-3.}\quad &
\end{array}
\right.
\end{eqnarray}
In order to search for the physical existence of the hyperscaling violated Yang-Mills-dilaton black hole, we investigate the thermal stability of this black hole.
\subsection{Thermal stability}
\begin{figure}
\centering
\subfigure[$det(H)$]{\includegraphics[scale=0.27]{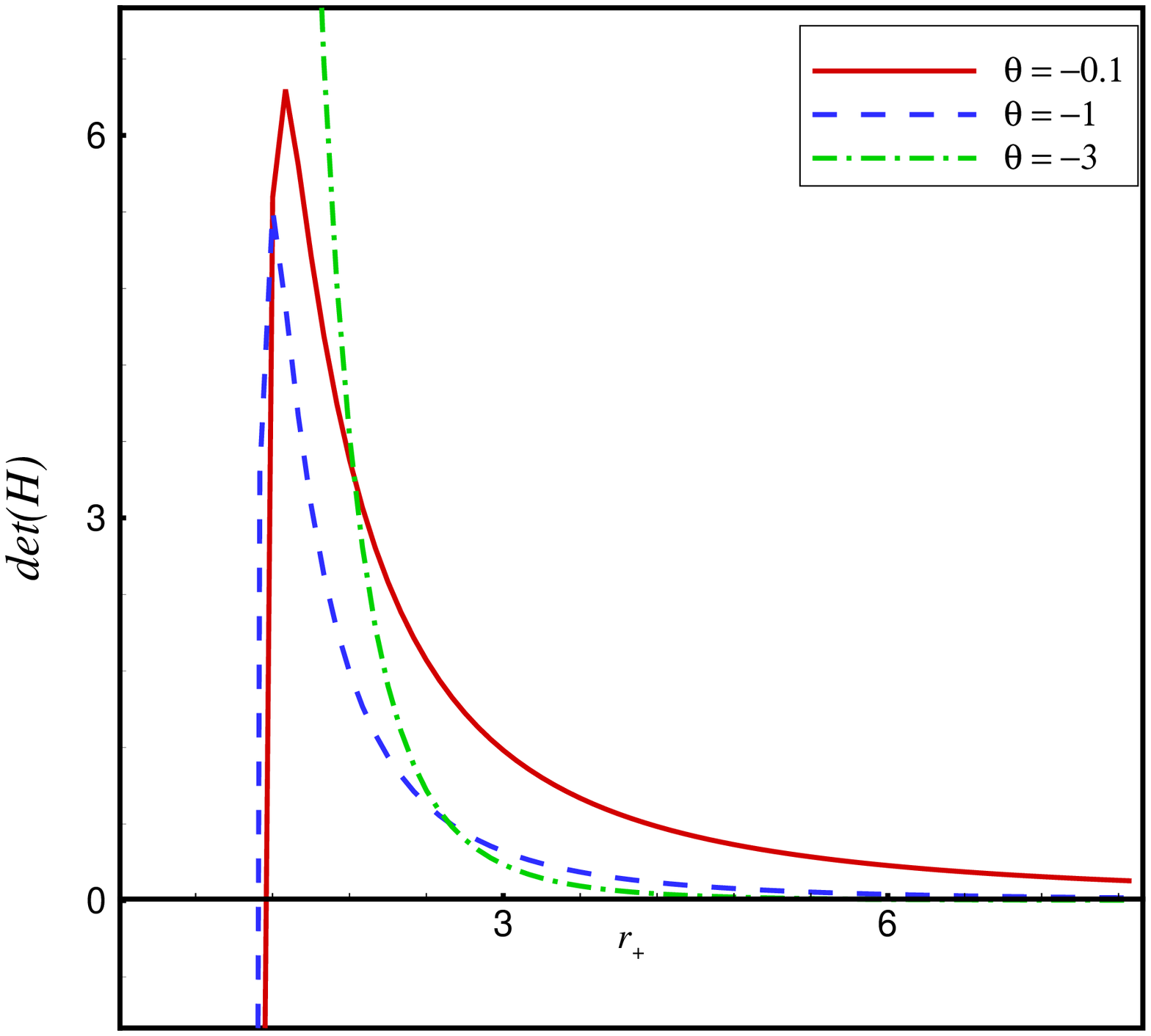}\label{Fig7a}}\hspace*{.2cm}
\subfigure[$d^2 M/dS^2$ ]{\includegraphics[scale=0.27]{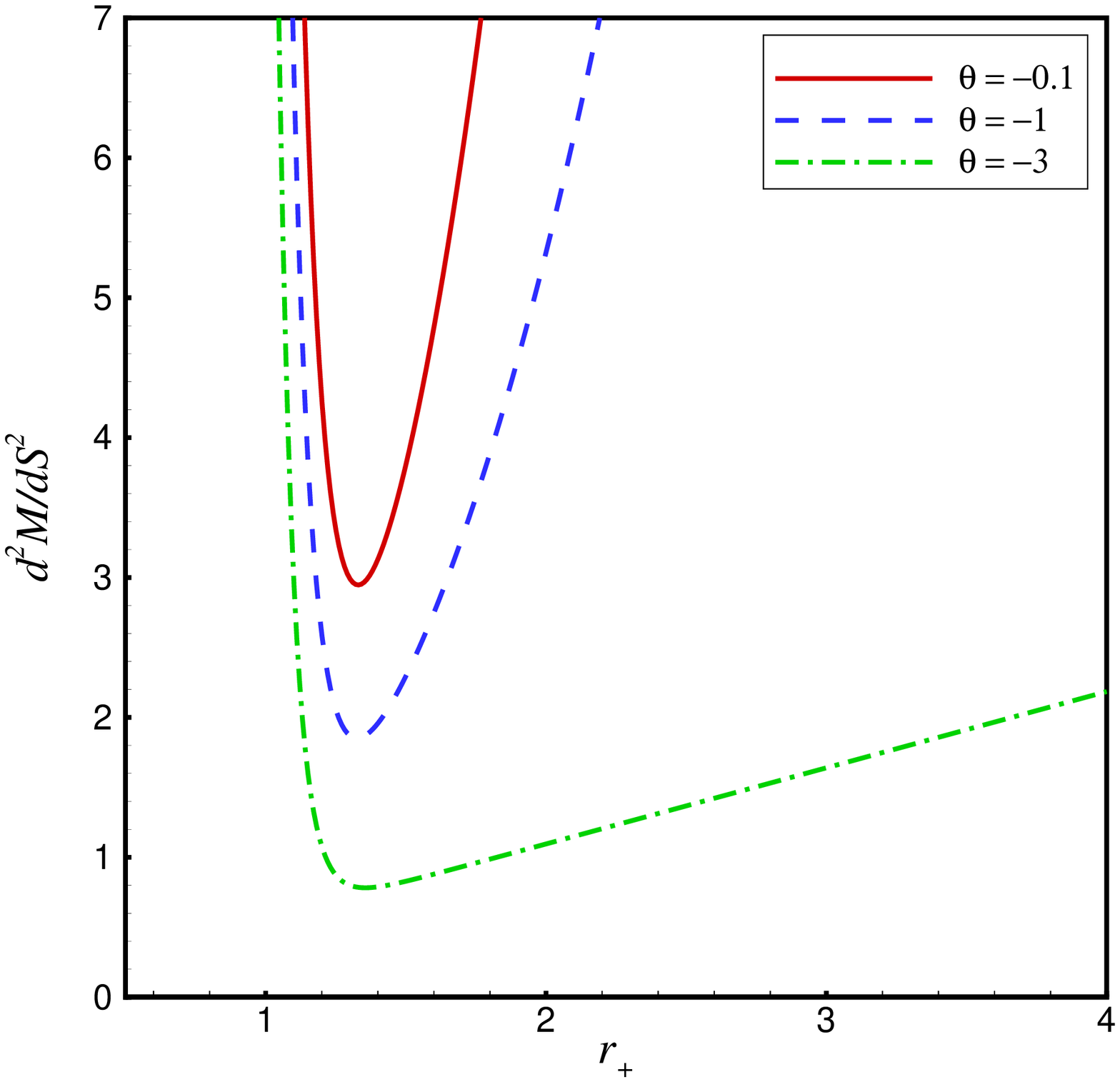}\label{Fig7b}}\hspace*{.2cm}
\subfigure[Temperature$T_{+}$]{\includegraphics[scale=0.27]{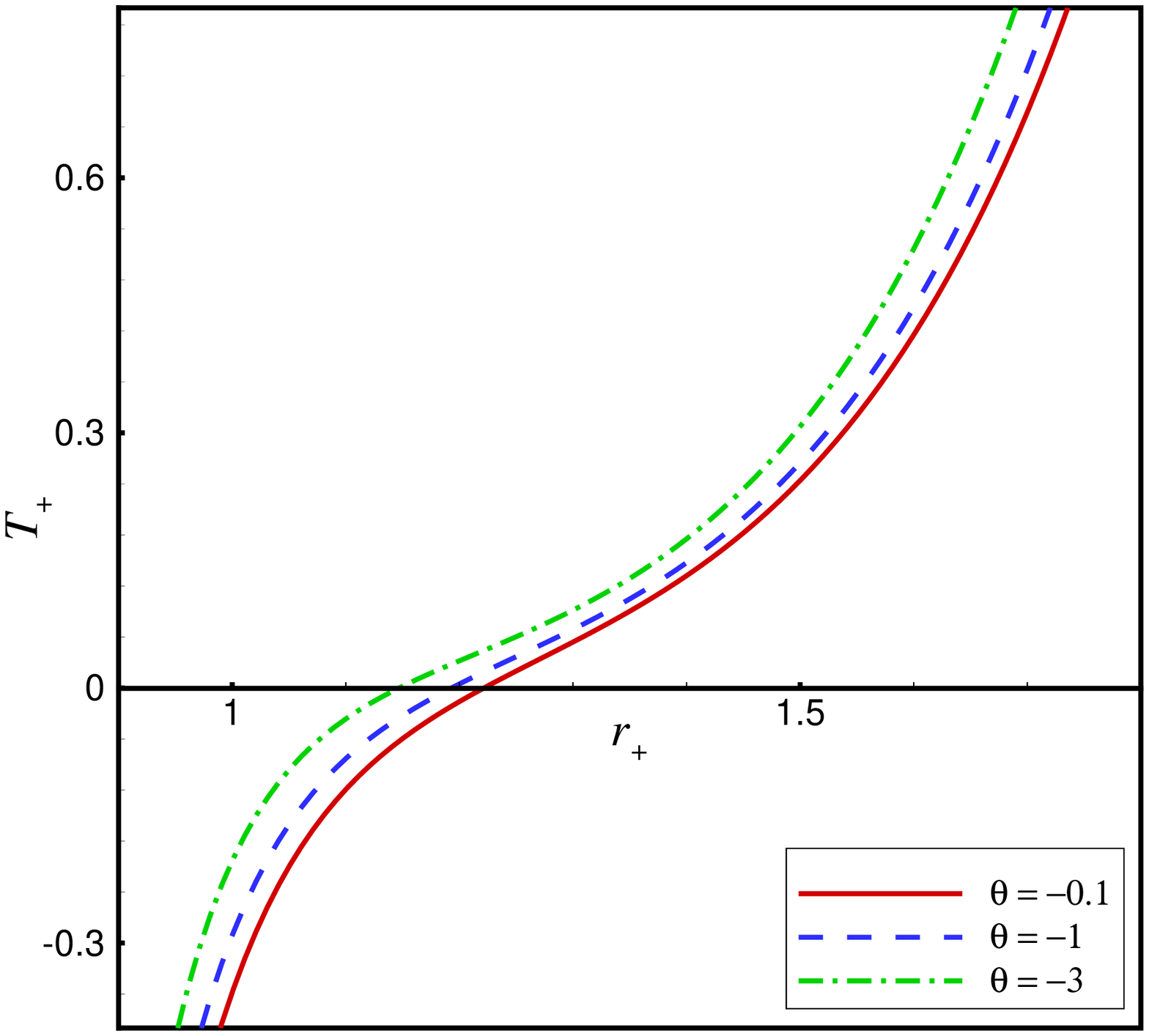}\label{Fig7c}}\caption{Thermal stability with respect to $r_{+}$ for different $\theta$ with $Q=1$, $z=8$ and $n=5$.}\label{Fig7}
\end{figure}

In the attempt of probing the thermal stability of the hyperscaling violated Yang-Mills-dilaton black hole, we follow the formalism in Sec.\ref{thermal1}. Therefore, we can obtain the Hessian matrix determinant of this black hole [we abbreviate it to $Det(H)$] as
\begin{eqnarray}\label{DetH1}
DetH&=&-\frac{2(n-1)^2(n-2)^2b^{2z-2-\frac{2\theta}{n-1}}r_{+}^{\frac{4\theta}{n-1}}}{[(n-1)(n-z-3)-\theta(n-5)](n-\theta-1)L^{4z-2}}\bigg(\frac{(n-1)(z-2)L^2}{[(n-1)z-2\theta]r_{+}^4}+\frac{z(z+n-\theta-1)}{(n-2)r_{+}^2}\nonumber\\
&&+\frac{16[(n-1)(2n-z-4)-2\theta(n-3)]\pi^2 Q^2b^{2z-2-\frac{2\theta}{n-1}}}{(n-\theta-1)L^{2z-4}r_{+}^{2z+4-\frac{4\theta}{n-1}}}\bigg), \,\,\,\,\,\,\,\,\,\,\,\,\,\,\,\,\,\,\,\,\,\,\,\,\, \mathrm{for}\,\, z\neq n-3-\frac{n-5}{n-1}\theta
\end{eqnarray}
\begin{eqnarray}\label{DetH2}
&&DetH=-\frac{64(n-1)^2(n-2)^2\pi^2 Q^2b^{4z-4-\frac{4\theta}{n-1}}}{(n-\theta-1)^2L^{6z-6}r_{+}^{2z+4-\frac{8\theta}{n-1}}}-\frac{2(n-1)(n-2)^2b^{2z-2-\frac{2\theta}{n-1}}r_{+}^{\frac{4\theta}{n-1}}\mathrm{ln}(r_{+})}{(n-\theta-1)L^{4z-2}}\bigg(\frac{(n-1)(z-2)L^2}{[z(n-1)-2\theta]r_{+}^4}+\nonumber\\
&&\frac{z(z+n-\theta-1)}{(n-2)r_{+}^2}+\frac{16[(n-1)(2n-z-4)-2\theta(n-3)]\pi^2 Q^2b^{2z-2-\frac{2\theta}{n-1}}}{(n-\theta-1)L^{2z-4}r_{+}^{2z+4-\frac{4\theta}{n-1}}}\bigg),\,\,\,\,\,\,\,\,\,\,\,\,\,\,\,\,\,\, \mathrm{for}\,\, z= n-3-\frac{n-5}{n-1}\theta.
\end{eqnarray}  
The thermal stability of the hyperscaling black hole may follow from a similar behavior as the Lifshitz one. For $z<n-3-\frac{n-5}{n-1}\theta$ and $z\geq2$, we face a negative value for $Det(H)$ and so there is an instability. On the other hand, if we select the solutions with $z>n-3-\frac{n-5}{n-1}\theta$, $Det(H)$ behaves differently for small and large $r_{+}$. For large $r_{+}$, the positive second term in the braces in Eq. \eqref{DetH1} is dominant and so $Det(H)$ is positive. For small $r_{+}$, it depends on the value of $(n-1)(2n-z-4)-2\theta(n-3)$ in the third term. To know more about this case, we have plotted $Det(H)$, $d^2 M/ dS^2$, and $T_{+}$ versus $r_{+}$ in Fig. \ref{Fig7}. It is clear from Fig. \ref{Fig7a} that, for large $r_{+}$, $Det(H)>0$, but for small $r_{+}$ with $n=5$ and $z=8$, $Det(H)$ depends on the values of $\theta$. For the mentioned parameters, $d^2M/dS^2>0$ and a positive temperature value recognizes the thermal stability. Our results show that the stable region usually happens for large $r_{+}$. \\
For $1\leq z<2$, $Det(H)$ is positive in dimensions $n=4,5$, independent of the values of $\theta$. So, $\theta$ cannot influence the stability regions of the solutions in the range $1\leq z<2$. Also, for $z=n-3-\frac{n-5}{n-1}\theta$, we cannot find a substantial stable region for the hyperscaling violated Yang-Mills-dilaton black hole.
\section{Critical behavior of the hyperscaling violated Yang-Mills-dilaton black hole}\label{PV2}
\begin{figure}
\centering
\subfigure[$Q^2-\Psi$ isotherms for $b=1$, $z=1.6$ and $n=3$ ]{\includegraphics[scale=0.27]{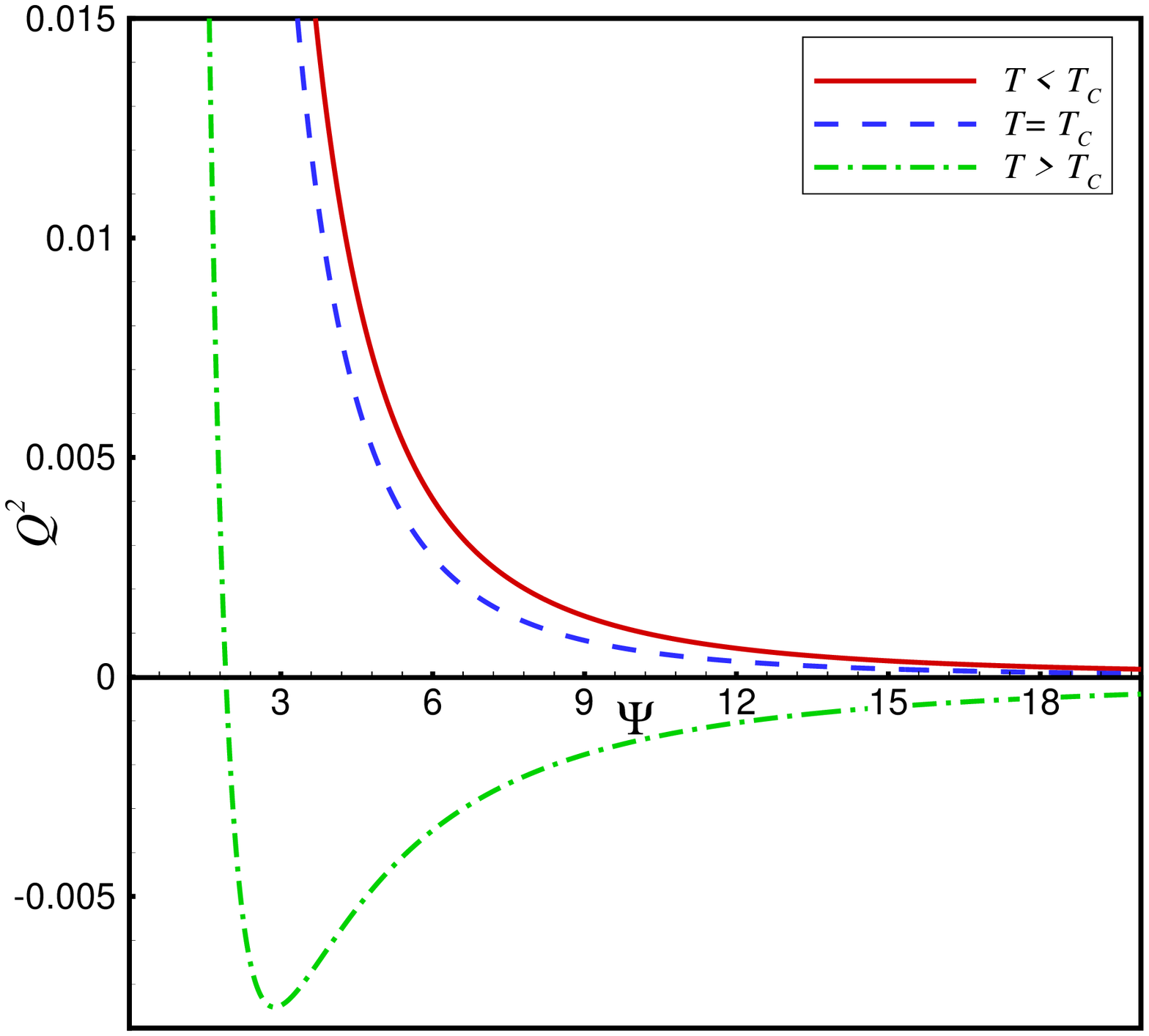}\label{Fig8a}}\hspace*{.2cm}
\subfigure[$G-Q^2$ diagrams for $b=2$, $z=1.2$ and $n=4$]{\includegraphics[scale=0.27]{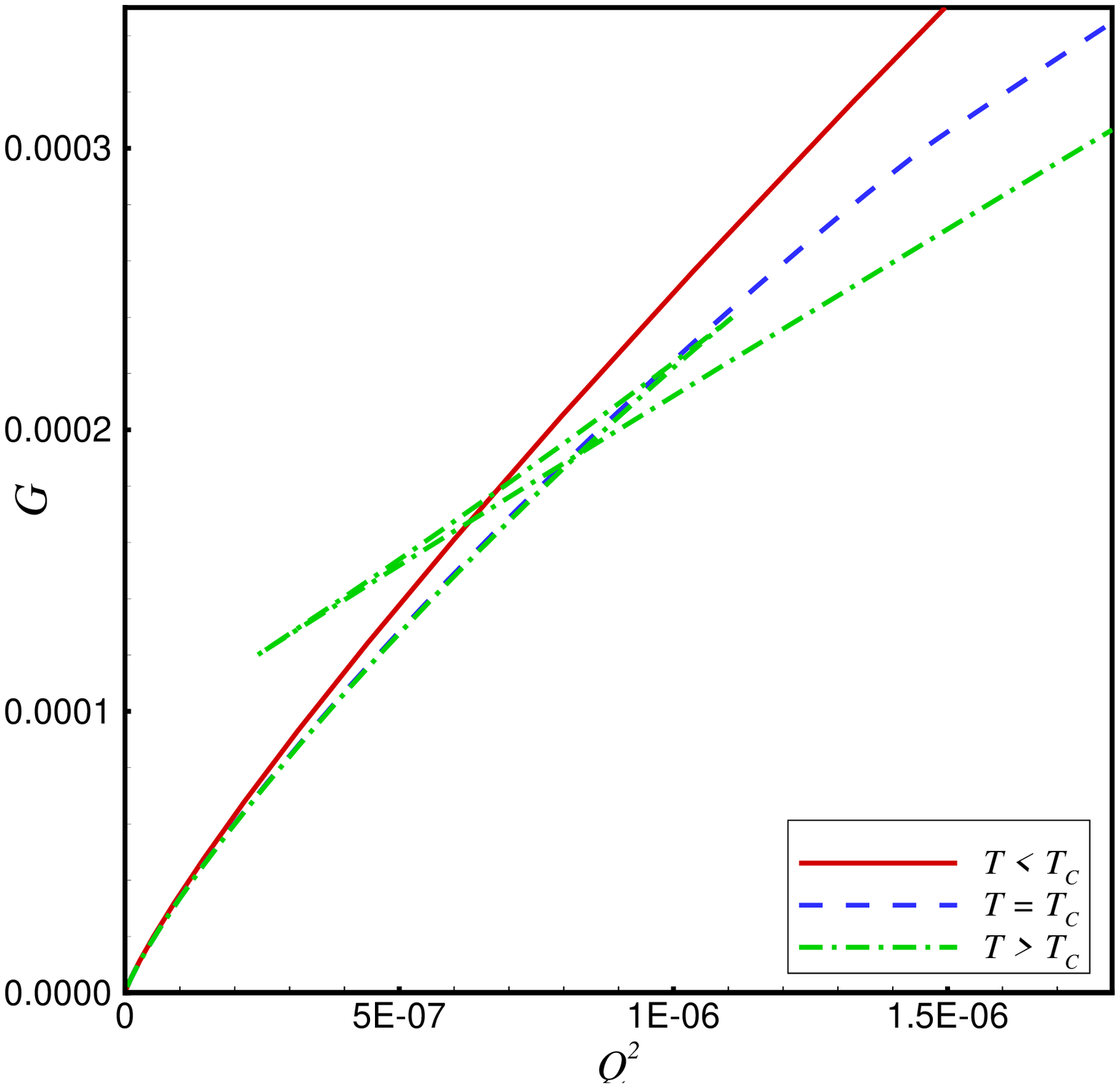}\label{Fig8b}}\caption{Critical behavior of the hyperscaling violated Yang-Mills-dilaton black hole with $\theta=-1$.}\label{Fig8}
\end{figure}

Now, we can use of the procedure in Sec. \ref{PV1} and probe the critical behavior of the hyperscaling violated Yang-Mills-dilaton black hole. In the extended phase space, the first law of thermodynamics obeys the rule \eqref{first}, where the volume is $V=r_{+}^{n-\theta}/(n-\theta)$ and
\begin{eqnarray}\label{Psi2}
\Psi=\bigg(\frac{\partial M}{\partial Q^2}\bigg)_{S,P}=-\frac{(n-1)^2(n-2)\pi b^{2z-2-\frac{2\theta}{n-1}}}{(n^2-n(\theta+z+4)+z+5\theta+3)L^{3z-3}}r_{+}^{-z+n-3-\frac{\theta(n-5)}{n-1}}.
\end{eqnarray}
If we consider the pressure as 
\begin{eqnarray}
P=\frac{(n-\theta)(n-\theta-1)r_{+}^{z-1}}{16\pi L^{z+1}},
\end{eqnarray}
the Smarr relation obeys from
\begin{eqnarray}
(z+n-\theta-3)M=(n-\theta-1)TS-2PV+\frac{2[(n-1)z-2\theta]}{n-1}\,\Psi Q^2\,\,\,\,\,\,\,\,\,\,\,\,\,\,\mathrm{for}\, z\neq n-3-\frac{n-5}{n-1}\theta.
\end{eqnarray}
Using Eq. \eqref{temp2}, we can obtain the equation of state for the hyperscaling violated black hole:
\begin{eqnarray}
Q^2&=&-\frac{(n-\theta-1)L^{3z-3}T}{4\pi(n-1)(n-2)b^{2z-2-\frac{2\theta}{n-1}}}r_{+}^{z+2-\frac{4\theta}{n-1}}+\frac{(n-\theta-1)L^{2z-2}}{16\pi^2 ((n-1)z-2\theta) b^{2z-2-\frac{2\theta}{n-1}}}r_{+}^{2z-\frac{4\theta}{n-1}}\nonumber\\
&&+\frac{(z+n-\theta-1)(n-\theta-1)L^{2z-4}}{16\pi^2(n-1)(n-2)b^{2z-2-\frac{2\theta}{n-1}}}r_{+}^{2z+2-\frac{4\theta}{n-1}},
\end{eqnarray} 
where $r_{+}$ is a function of $\Psi$ in Eq. \eqref{Psi2}. The Gibbs free energy of the hyperscaling violated Yang-Mills-dilaton black hole is obtained from the relation
\begin{eqnarray}
G&=&M-TS,
\end{eqnarray}
that the critical points of this black hole are read from  
\begin{eqnarray}
\Psi_{C}=-\frac{(n-1)^2(n-2)\pi b^{2(z-1)-\frac{2\theta}{n-1}}}{[(-z+n-3)(n-1)-\theta(n-5)]L^{3z-3}}\bigg(-\frac{(n-1)(n-2)(z-2)L^2}{z((z+1)(n-1)-2\theta)(z+n-\theta-1)}\bigg)^{\frac{-z+n-3}{2}-\frac{(n-5)\theta}{2(n-1)}},
\end{eqnarray}
\begin{eqnarray}\label{Tcriticalv}
T_{C}=-\frac{((n-1)(z+1)-2\theta)(z+n-\theta-1)}{(z-2)((n-1)(z+2)-4\theta)\pi L^{z+1}}\bigg(-\frac{(n-1)(n-2)(z-2)L^2}{z((n-1)(z+1)-2\theta)(z+n-\theta-1)}\bigg)^{\frac{z}{2}}.
\end{eqnarray}
We have plotted the $Q^2-\Psi$ isotherms and $G-Q^2$ diagrams of this black hole in Fig. \ref{Fig8}. The isotherms behave like the van der Waals gas and the Gibbs free energy shows a first-order phase transition from a black hole with small $r_{+}$ to a one with large $r_{+}$. 
\section{concluding remark}\label{result} 
In this paper, we achieved two new sets of higher-dimensional Lifshitz and hyperscaling violated dilaton black holes solutions in the presence of the Yang-Mills gauge fields. Using the Wu-Yang ansatz, we considered three Yang-Mills gauge fields, each with $SO(n)$ and $SO(n-1,1)$ gauge symmetric groups. In the first part of this paper, we paid attention to the Lifshitz spacetime with a dynamical exponent $z$ and obtained the Lifshitz-Yang-Mills-dilaton black hole solutions. In the second part, we used a spacetime that, in addition to the parameter $z$, also has a hyperscaling violation parameter $\theta$. For these solutions, we considered a potential with a Liouville form for the dilaton field. The obtained results manifested that, in order to have the asymptotic real solutions, we must fix $\theta<0$. \\
The obtained field equations led to the real solutions only for the spherical hypersurface in the AdS Lifshitz and hyperscaling violated spacetimes. We also had to fix the coupling constants of the two gauge fields to support the Lifshitz and hyperscaling violated spacetimes. \\
For $n>3$, the obtained Yang-Mills-dilaton black hole solutions are different from the ones in the Maxwell theory. The Maxwell-dilaton Lifshitz solutions have just one special form for each value of $z$, while our obtained Lifshitz-Yang-Mills-dilaton solutions are set in two categories with $z\neq n-3$ and $z=n-3$. This is also true for the obtained hyperscaling violated Yang-Mills-dilaton solutions, and they are divided into two parts, $z\neq n-3-\frac{n-5}{n-1}\theta$ and $z=n-3-\frac{n-5}{n-1}\theta$. It is clear that $\theta=0$ implies the Lifshitz solutions.\\
We also studied the physical structures of the black holes. The solutions announce an essential singularity at the origin. Depending on the values of the parameters $m$, $e_{3}$, $z$, $n$, and $\theta$, a black hole with two horizons, an extreme black hole, or a naked singularity may form from the obtained solutions. For the large values of the mass parameter $m$, it is more possible to have a black hole with inner and outer horizons.\\
We also studied the thermodynamic behaviors of the black hole solutions. We obtained the thermodynamic quantities such as mass, temperature, entropy, charge, and electric potential which satisfy the first law of thermodynamics. We also probed the thermal stability of the obtained solutions in the grand canonical ensemble. For this purpose, the positive values of the quantities such as the Hessian matrix determinant, $\big(\partial ^2 M/\partial S^2\big)_{Q}$, and temperature are necessary. 
The hyperscaling violated Yang-Mills-dilaton black hole solutions with $z\leq n-3-\frac{n-5}{n-1}\theta$ and $z\geq 2$ are not stable, while for $z>n-3-\frac{n-5}{n-1}\theta$, the stability depends on the values of $(n-1)(2n-z-4)-2\theta(n-3)$ and $T_{+}$; i.e., if $(n-1)(2n-z-4)-2\theta(n-3)$ is positive, $Det(H)$(Hessian matrix determinant for the hyperscaling violated black holes) is positive for all $r_{+}$, and so a positive temperature determines stability. However, if $(n-1)(2n-z-4)-2\theta(n-3)$ is negative, $Det(H)$ has a negative value for small $r_{+}$, and so we should find a unit positive region between $Det(H)$ and $T_{+}$. For $1\leq z<2$, $Det(H)$ is positive only for $n=3,4$, and so the stability happens in these dimensions, if $T_{+}>0$. The thermal stability of the Lifshitz-Yang-Mills-dilaton black hole obeys from a similar behavior, if we choose the condition $\theta=0$. If we choose a large value for the charge $Q$, then the Lifshitz black hole includes a large region in thermal stability.\\
We also checked out the critical behavior of both black holes. We used a method in which both the cosmological constant $\Lambda$ and the square charge $Q^2$ play as thermodynamic variables and the pressure $P$ and $\Psi=\partial M/\partial Q^2$ are the related conjugates. We could get to a Smarr-type formula for $z\neq n-3$ and $z\neq n-3-\frac{n-5}{n-1}\theta$ in Lifshitz and hyperscaling violated spacetimes, respectively. For the obtained solutions with $1\leq z< 2$, we found a similar behavior between the isotherms of the black holes ($Q^2-\Psi$ isotherms) and the ones in van der Waals gas. Also, the obtained results indicated a small-large black hole phase transition at the critical point for $T>T_{C}$.\\ 
In the future, we intend to use these new non-Abelian Yang-Mills Lifshitz solutions to obtain useful information about possible dual systems such as ferromagnet spin currents and quark confinement.
We may also extend this study and investigate the solutions for the other gauge groups. It is also possible to probe the Joule-Thomson expansion and obtain the quasinormal modes of the Lifshitz and hyperscaling violated Yang-Mills-dilaton black holes.
\section{APPENDIX: Details of some gauge groups} \label{app}
To know more, here are the details of some gauge groups. For the four-dimensional spacetime($n=3$), the metric is defined as below: 
\begin{eqnarray}
ds^2=-\frac{r^{2z}}{L^{2z}}f(r)dt^2+\frac{L^2}{r^2f(r)}dr^2+r^2 d\theta^2+r^2\left\{
\begin{array}{ll}
$$\mathrm{sin}^{2}\,\theta\,d\phi^2 $$,\quad\quad\quad \quad  \ {\mathrm{for}\,\, k=1, }\quad &  \\ \\
$$\mathrm{sinh}^{2}\,\theta\,d\phi^2 $$,\quad\quad\quad  \ {\mathrm{for}\,\, k=-1.}\quad &
\end{array}
\right.
\end{eqnarray} 
If we consider the coordinates \eqref{coor} for $k=1$ and $n=3$,
\begin{eqnarray}
x_{1}&=&r\, \mathrm{sin}\,\theta\, \mathrm{cos}\, \phi,\nonumber\\
x_{2}&=&r \,\mathrm{sin}\,\theta\, \mathrm{sin}\, \phi,\nonumber\\
x_{3}&=&r\, \mathrm{cos}\,\theta,\nonumber\\
\end{eqnarray}
then the gauge potentials \eqref{poten} of the gauge group $SO(3)$ are obtained as 
\begin{eqnarray}
A^{(1)}&=&\frac{e}{r^2}(x_{1}dx_{3}-x_{3}dx_{1}),\nonumber\\
A^{(2)}&=&\frac{e}{r^2}(x_{2}dx_{3}-x_{3}dx_{2}),\nonumber\\
A^{(3)}&=&\frac{e}{r^2}(x_{1}dx_{2}-x_{2}dx_{1}).\nonumber\\
\end{eqnarray}
By simplification, they reduce to 
\begin{eqnarray}
A_{\mu}^{(1)}&=& e\,(-\mathrm{cos}\, \phi\, d\theta+\mathrm{sin}\, \theta \,\mathrm{cos}\,\theta \,\mathrm{sin}\,\phi \,d\phi),\nonumber\\
A_{\mu}^{(2)}&=& -e\,(\mathrm{sin}\, \phi\, d\theta+\mathrm{sin}\, \theta \,\mathrm{cos}\,\theta \,\mathrm{cos}\,\phi \,d\phi),\nonumber\\
A_{\mu}^{(3)}&=& e\,\mathrm{sin}^{2}\, \theta \,d\phi,
\end{eqnarray}
where the coupling constants are $C^{1}_{23}= C^{2}_{31}=C^{3}_{12}=-1$ and $\gamma_{ab}=\mathrm{diag}(1,1,1)$. 
We also define the gauge potentials of the $SO(2,1)$ gauge group in the four-dimensional spacetime with $k=-1$ as follows:
\begin{eqnarray}
A_{\mu}^{(1)}&=& e\,(-\mathrm{cos}\, \phi\, d\theta+\mathrm{sinh}\, \theta \,\mathrm{cosh}\,\theta \,\mathrm{sin}\,\phi \,d\phi),\nonumber\\
A_{\mu}^{(2)}&=& -e\,(\mathrm{sin}\, \phi\, d\theta+\mathrm{sinh}\, \theta \,\mathrm{cosh}\,\theta \,\mathrm{cos}\,\phi \,d\phi),\nonumber\\
A_{\mu}^{(3)}&=& e\,\mathrm{sinh}^{2}\, \theta \,d\phi,
\end{eqnarray}
where $C^{1}_{23}= C^{2}_{31}=-C^{3}_{12}=1$ and $\gamma_{ab}=\mathrm{diag}(-1,-1,1)$. \\
If we choose the five-dimensional spacetime ($n=4$), 
\begin{eqnarray}
ds^2=-\frac{r^{2z}}{L^{2z}}f(r)dt^2+\frac{L^2}{r^2f(r)}dr^2+r^2 d\theta^2+r^2\left\{
\begin{array}{ll}
$$\mathrm{sin}^{2}\,\theta\,( d\phi^2+\mathrm{sin}^{2}\,\phi\,d\psi^2) $$,\quad\quad\quad \quad  \ {\mathrm{for}\,\, k=1, }\quad &  \\ \\
$$\mathrm{sinh}^{2}\,\theta\,( d\phi^2+\mathrm{sin}^{2}\,\phi\, d\psi^2)$$,\quad\quad\quad  \ {\mathrm{for}\,\, k=-1,}\quad &
\end{array}
\right.
\end{eqnarray} 
then the gauge potentials of the $SO(4)$ gauge group with $k=1$ are 
\begin{eqnarray}
A_{\mu}^{(1)}&=& -e\,(\mathrm{sin}\, \phi\,\mathrm{cos}\, \psi\, d\theta+\mathrm{sin}\, \theta \,\mathrm{cos}\,\theta \,(\mathrm{cos}\,\phi\, \mathrm{cos}\,\psi \,d\phi-\mathrm{sin}\,\phi\, \mathrm{sin}\,\psi \,d\psi)),\nonumber\\
A_{\mu}^{(2)}&=& -e\,(\mathrm{sin}\, \phi\,\mathrm{sin}\, \psi\, d\theta+\mathrm{sin}\, \theta \,\mathrm{cos}\,\theta \,(\mathrm{cos}\,\phi\, \mathrm{sin}\,\psi \,d\phi+\mathrm{sin}\,\phi\, \mathrm{cos}\,\psi \,d\psi)),\nonumber\\
A_{\mu}^{(3)}&=& -e\,(\mathrm{cos}\, \phi\, d\theta-\mathrm{sin}\, \theta \,\mathrm{cos}\,\theta \,\mathrm{sin}\,\phi \,d\phi),\nonumber\\
A_{\mu}^{(4)}&=& -e\,\mathrm{sin}^{2}\, \theta \,\mathrm{sin}^{2}\, \phi \,d\psi,\nonumber\\
A_{\mu}^{(5)}&=& e\,\mathrm{sin}^{2}\, \theta\,(\mathrm{cos}\, \psi\, d\phi-\mathrm{sin}\, \phi \,\mathrm{cos}\,\phi \,\mathrm{sin}\,\psi \,d\psi),\\
A_{\mu}^{(6)}&=& e\,\mathrm{sin}^{2}\, \theta\,(\mathrm{sin}\, \psi\, d\phi+\mathrm{sin}\, \phi \,\mathrm{cos}\,\phi \,\mathrm{cos}\,\psi \,d\psi),
\end{eqnarray}
where the coupling constants are defined
\begin{eqnarray}
C^{1}_{24}= C^{1}_{35}=C^{2}_{41}=C^{2}_{36}=C^{3}_{51}=C^{3}_{62}=1\,,\,
C^{4}_{56}= -C^{4}_{21}=C^{5}_{64}=-C^{5}_{31}=C^{6}_{45}=-C^{6}_{32}=1,
\end{eqnarray}
and $\gamma_{ab}=\mathrm{diag}(1,1,1,1,1,1)$.\\
For the gauge group $SO(3,1)$ with $k=-1$, the gauge potentials have the forms
\begin{eqnarray}
A_{\mu}^{(1)}&=& -e\,(\mathrm{sin}\, \phi\,\mathrm{cos}\, \psi\, d\theta+\mathrm{sinh}\, \theta \,\mathrm{cosh}\,\theta \,(\mathrm{cos}\,\phi\, \mathrm{cos}\,\psi \,d\phi-\mathrm{sin}\,\phi\, \mathrm{sin}\,\psi \,d\psi)),\nonumber\\
A_{\mu}^{(2)}&=& -e\,(\mathrm{sin}\, \phi\,\mathrm{sin}\, \psi\, d\theta+\mathrm{sinh}\, \theta \,\mathrm{cosh}\,\theta \,(\mathrm{cos}\,\phi\, \mathrm{sin}\,\psi \,d\phi+\mathrm{sin}\,\phi\, \mathrm{cos}\,\psi \,d\psi)),\nonumber\\
A_{\mu}^{(3)}&=& -e\,(\mathrm{cos}\, \phi\, d\theta-\mathrm{sinh}\, \theta \,\mathrm{cosh}\,\theta \,\mathrm{sin}\,\phi \,d\phi),\nonumber\\
A_{\mu}^{(4)}&=& e\,\mathrm{sinh}^{2}\, \theta \,\mathrm{sin}^{2}\, \phi \,d\psi,\nonumber\\
A_{\mu}^{(5)}&=& -e\,\mathrm{sinh}^{2}\, \theta\,(\mathrm{cos}\, \psi\, d\phi-\mathrm{sin}\, \phi \,\mathrm{cos}\,\phi \,\mathrm{sin}\,\psi \,d\psi),\\
A_{\mu}^{(6)}&=& -e\,\mathrm{sinh}^{2}\, \theta\,(\mathrm{sin}\, \psi\, d\phi+\mathrm{sin}\, \phi \,\mathrm{cos}\,\phi \,\mathrm{cos}\,\psi \,d\psi).
\end{eqnarray}
where $C^{1}_{24}= C^{1}_{35}=C^{2}_{41}=C^{2}_{36}=C^{3}_{51}=C^{3}_{62}=1$, $C^{4}_{56}= C^{4}_{21}=C^{5}_{64}=C^{5}_{31}=C^{6}_{45}=C^{6}_{32}=1$, and\,\,\,\,\,\,\,\,\,\,\,\,\,\,\,\,\,\,\,\,\,\,\,\,\,\,\,\,\,\,\,\,\,\,\,\,\,\,\,\,\,\,\,\,\,\,\,\,\,\,\,\,\,\,\, $\gamma_{ab}=\mathrm{diag}(-1,-1,-1,1,1,1)$.

\acknowledgments{This work is supported by Irainian National Science Foundation (INSF). F.N would like to thank physics department of Isfahan University of Technology for warm hospitality.}

\end{document}